\documentclass[longtitle,preprint,5p,times,twocolumn]{elsarticle}

\usepackage{amssymb}
\usepackage{amsmath}

\usepackage{caption}
\usepackage{xspace}
\usepackage{longtable}
\usepackage{pdflscape}
\usepackage{multirow}
\usepackage{ragged2e}
\usepackage{booktabs}
\usepackage{hyperref}

\usepackage[table]{xcolor}
\usepackage{comment}
\usepackage{lipsum}

\newcommand{\geant}{\mbox{\sc Geant4}\xspace}

\usepackage{lineno}

\usepackage{mathtools}

\DeclarePairedDelimiter\floor{\lfloor}{\rfloor}

\usepackage{array}
\newcolumntype{P}[1]{>{\centering\arraybackslash}p{#1}}
\newcolumntype{M}[1]{>{\centering\arraybackslash}m{#1}}

\usepackage{mfirstuc} % to uppercase the name

\newcommand{\addComment}[2]{
  \expandafter\newcommand\csname #1\endcsname[1]{{\bf \color{#2} \capitalisewords{#1}:\,##1}}
  \expandafter\newcommand\csname #1cor\endcsname[2]{{\color{#2} \capitalisewords{#1}:\,\st{##1}{\bf ##2}}}
  \expandafter\newcommand\csname #1color\endcsname{#2}
}

\usepackage[normalem]{ulem}

\usepackage{adjustbox}
\usepackage{gensymb}

\usepackage{multirow}
\usepackage{multicol}

\addComment{cris}{blue} 
\addComment{karthik}{red} 
\journal{Nuclear Instruments and Methods A}

\begin{document}

\begin{frontmatter}

\title{AI-assisted Optimization of the ECCE Tracking System at the Electron Ion Collider}
%\title{AI-assisted Design of the ECCE Detector: the Tracking System}
%\title{AI-assisted Detector Design at EIC: \\the ECCE Tracker Example}

% I will put these into the new NIM format.
% See the example.   Author list is fine, just the affliation list in their nice bibtex like format.

%add "longtitle" as elsarticle option above to allow page break with long list of institutes - ok DWH

% Just leaving the one here and will upload the corrected one later.  
% And add "longtitle" got it.

\def\theaffn{\arabic{affn}}
%
% Full list of ECCE Collaboratotions
%
% Click Refresh to get the latest version
%
% Goto the following to edit:
% https://www.overleaf.com/1234763137gmqgnzsrpsdj

\author[MIT]{C.~Fanelli}
\author[Regina]{Z.~Papandreou}
\author[Regina]{K.~Suresh}
\author[MoreheadState]{J.~K.~Adkins}
\author[RIKEN,RBRC]{Y.~Akiba}
\author[UKansas]{A.~Albataineh}
\author[ODU]{M.~Amaryan}
\author[Oslo]{I.~C.~Arsene}
\author[MSU]{C. Ayerbe Gayoso}
\author[Sungkyunkwan]{J.~Bae}
% ...

%\input{authors/authorlist_new}
%\input{authors/affiliations-test}
%
% Full ECCE author list
% Click Refresh to get the latest version
%
% Goto the following to edit:
% https://www.overleaf.com/1234763137gmqgnzsrpsdj

%
% Please fill in any missing information
%
%
% Example of the format:
%
% \affiliation[]{organization={},
%	 addressline={},
%	 city={},
%	 postcode={},
%	 state={},
%	 country={}} 
%

% ...
\affiliation[MoreheadState]{organization={Morehead State University},
	 city={Morehead},
	 postcode={},
	 state={KY},
	 }
 
\affiliation[MIT]{organization={Massachusetts Institute of Technology},
	 addressline={77 Massachusetts Ave.},
	 city={Cambridge},
	 postcode={02139},
	 state={MA},
	 country={USA}} 
 
\affiliation[MSU]{organization={Mississippi State University},
	 city={Mississippi State},
	 postcode={},
	 state={MS},
	 country={USA}} 
 
% ...
\affiliation[ODU]{organization={Old Dominion University},
	 city={Norfolk},
	 postcode={},
	 state={VA},
	 country={USA}} 
 
% ...
\affiliation[RIKEN]{organization={RIKEN Nishina Center},
	 city={Wako},
	 state={Saitama},
	 country={Japan}} 
 
% ...
\affiliation[RBRC]{organization={RIKEN BNL Research Center},
	 city={Upton},
	 postcode={11973},
	 state={NY},
	 country={USA}} 
% ...
\affiliation[Sungkyunkwan]{organization={Sungkyunkwan University},
	 city={Suwon},
	 country={Republic of Korea}} 
 
% ...
\affiliation[UKansas]{organization={Unviersity of Kansas},
	 addressline={1450 Jayhawk Blvd.},
	 city={Lawrence},
	 postcode={66045},
	 state={KS},
	 country={USA}} 
 
% ...
\affiliation[Oslo]{organization={University of Oslo},
	 city={Oslo},
	 country={Norway}} 
 
\affiliation[Regina]{organization={ University of Regina},
	 city={Regina},
	 postcode={},
	 state={SK},
	 country={Canada}} 
 
% ...

%\begin{abstract}
%\end{abstract}
%\linenumbers 

\begin{abstract}
The Electron-Ion Collider (EIC) is a cutting-edge accelerator facility that will study the nature of the ``glue'' that binds the building blocks of the visible matter in the universe. The proposed experiment will be realized at Brookhaven National Laboratory in approximately 10 years from now, with detector design and R\&D currently ongoing. 
Notably, EIC is one of the first large-scale facilities to leverage Artificial Intelligence (AI) %during the design and R\&D phases. 
 already starting from the design and R\&D phases. 
%The EIC Comprehensive Chromodynamics Experiment (ECCE) is a consortium that is proposing a detector design based on a 1.5T solenoid. Herein we describe a comprehensive optimization of the ECCE tracker using AI. 
The EIC Comprehensive Chromodynamics Experiment (ECCE) is a consortium that proposed a detector design based on a 1.5T solenoid. The EIC detector proposal review concluded that the ECCE design will serve as the reference design for an EIC detector. 
Herein we describe a comprehensive optimization of the ECCE tracker using AI. 
The work required a complex parametrization of the simulated detector system. % simulated with the Fun4All framework.  
Our approach dealt with an optimization problem in a multidimensional design space driven by multiple objectives that encode the detector performance, while satisfying several mechanical constraints. 
We describe our strategy and show results obtained for the ECCE tracking system. 
The AI-assisted design is agnostic to the simulation framework and can be extended to other sub-detectors or to a system of sub-detectors to further optimize the performance of the EIC detector.

\end{abstract}


\end{frontmatter}

\setcounter{tocdepth}{1}
\tableofcontents

%% 
%% Input Sections of the Document
%%
%\input{sections/overview}
%\input{sections/perform}
%\input{sections/summary}
%\linenumbers
%%%%%%%%%%%%%%%%%%%%%%%%%%%%%%%%%%%%%%%%%%%%%%%%%
\section{Introduction}\label{sec:intro}
%%%%%%%%%%%%%%%%%%%%%%%%%%%%%%%%%%%%%%%%%%%%%%%%%

The Electron Ion Collider (EIC) \cite{khalek2021science} is a future cutting-edge discovery machine that will unlock the secrets of the gluonic force binding the building blocks of the visibile matter in the universe. The EIC will consist of two intersecting accelerators, one producing an intense beam of electrons and the other a beam of protons or heavier atomic nuclei; it will be the only electron-nucleus collider operating in the world. 
The EIC Comprehensive Chromodynamics Experiment (ECCE) \cite{ecce} is an international consortium assembled to develop a detector that can offer full energy coverage and an optimized far forward detection region.
ECCE has investigated a detector design based on the existing BABAR 1.5T magnet; %solenoid in both EIC interaction points %(namely IP6 and IP8 with 25 and $\sim$35 mrad crossing angle, respectively),
 this detector will be ready for the beginning of EIC operations. 
 More details on the ECCE detector design and what is described in the following can be found in \cite{ecce-paper-det-2022-01}. 

ECCE is an integrated detector that extends for about 40~m, and includes a central detector built around the interaction point and far-forward (hadron-going direction) and far-backward (electron-going direction) regions \cite{khalek2021science}. To fulfill the physics goals of the EIC, the central detector needs to be hermetic and provide good particle identification (PID) over a large phase space. 
The central detector itself consists of multiple sub-detectors: a tracking system made by inner and outer tracker stations allows the reconstruction of charged particles moving in the magnetic field; a system of PID sub-detectors will cover the barrel and the electron-going and hadron-going directions; electromagnetic and hadronic calorimeters are used to detect showers and provide complete information on the particle flow which is essential for certain event topologies, \textit{e.g.}, those containing jets. 
%
%The optimization of the EIC detector is of utmost importance during the design and R\&D phases \cite{ent_eic_schedule_2021}.

As outlined in \cite{khalek2021science}, Artificial Intelligence (AI) can provide dedicated strategies for complex combinatorial searches and can handle multi-objective problems characterized by a multidimensional design space, allowing the identification of hidden correlations among the design parameters. 
ECCE included these techniques in the design workflow during the detector proposal. 
At first this AI-assisted design strategy was used to steer the design. 
%
%After the technology choice is selected (based also on the important insights provided by AI), it can be further utilized to eventually fine-tune the detector parameters. 
After the base technology is selected using insights provided by AI, its detector parameters can be further fine-tuned using AI.  
During the ECCE detector proposal stage, the design of the detector underwent a continual optimization process \cite{ent_eic_schedule_2021}. % that evolved in time. %
 %In the following we describe the role played by the AI-assisted workflow and highlight this as a novel aspect introduced during the design process of an experiment.  

%This technical note is therefore structured as follows: in Sec. \ref{sec:ai_assisted} we describe the implemented AI pipeline that supported the design optimization in a sequential strategy the fosters the interplay between the different working groups in a post hoc decision making process; in Sec. \ref{conclusions} we will present our conclusions and provide a summary with future directions. 

The article is structured as follows: in Sec.~\ref{sec:ai_assisted} we provide an overview of design optimization and describe the AI-assisted strategy; 
in Sec.~\ref{sec:simulation} we introduce the ECCE tracker and describe the software stack utilized in this work to which AI is coupled for the optimization;
in Sec.~\ref{sec:analysis} we describe the implemented pipeline that results in a sequential strategy, fostering the interplay between the different working groups in a post hoc decision making process; 
in Sec.~\ref{sec:computing} we present perspectives and planned activities. 

The ECCE detector at the EIC will be one of the first examples of detectors that will be realized leveraging AI during the design and R\&D phases.

%Detector optimization is an essential part of the R\&D and design process that involves mechanical design and budget to realize the best performance possible. This process is anticipated to continue in the months following the detector proposal towards CD-2 and CD-3.

%%%%%%%%%%%%%%%%%%%%%%%%%%%%%%%%%%%%%%%%%%%%%%%%%
\section{AI-assisted Detector Design}\label{sec:ai_assisted}
%%%%%%%%%%%%%%%%%%%%%%%%%%%%%%%%%%%%%%%%%%%%%%%%%

Detector optimization with AI is 
%an essential part of the ECCE R\&D and design process and it is 
anticipated to continue in the months following the detector proposal towards CD-2 and CD-3. %according to the official EIC schedule \cite{ent_eic_schedule_2021}. 
Optimizing the design of large-scale detectors such as ECCE---that are made of multiple sub-detector systems---is a complex problem. 
Each sub-detector system is characterized by a multi-dimensional design parameter space. 
In addition, detector simulations are typically computationally intensive, and rely on advanced simulation platforms used in our community such as \geant \cite{agostinelli2003geant4} to simulate the interaction of radiation with matter.
Additional computationally expensive steps are present along the  data reconstruction and analysis pipeline.  
The software stack that is utilized in the detector design process involves three main steps:
(i) generation of events, (ii) detector simulations
and (iii) reconstruction and analysis.

%Another level of complexity to consider in detector design is the reconstruction of features from the detector response and eventually physics events analysis which allows to build objectives that drive the detector design.  

%%of complexity when it comes to detector design is added by the reconstruction and analysis of events, where certain tasks like pattern recognition for particle identification can be computationally expensive.

As pointed out in \cite{fanelli2022design}, the above bottlenecks render the generation and exploration of mutliple design points cumbersome. 
This in turn represents an obstacle for deep learning (DL)-based approaches that learn the mapping between the design space and the functional space \cite{zhou2019optimization,guo2021artificial,sanchez2018inverse}, which could facilitate the identification of optimal design points. 
In principle fast simulations with DL can reduce the most CPU-intensive parts of the simulation and provide accurate results \cite{atlas2021atlfast3}, although   several design points need to be produced with \geant before injection in any DL architecture. Similar considerations exist in deploying DL for reconstruction during the design optimization process. 

\begin{figure}[!]
    \centering
    \includegraphics[width=0.45\textwidth]{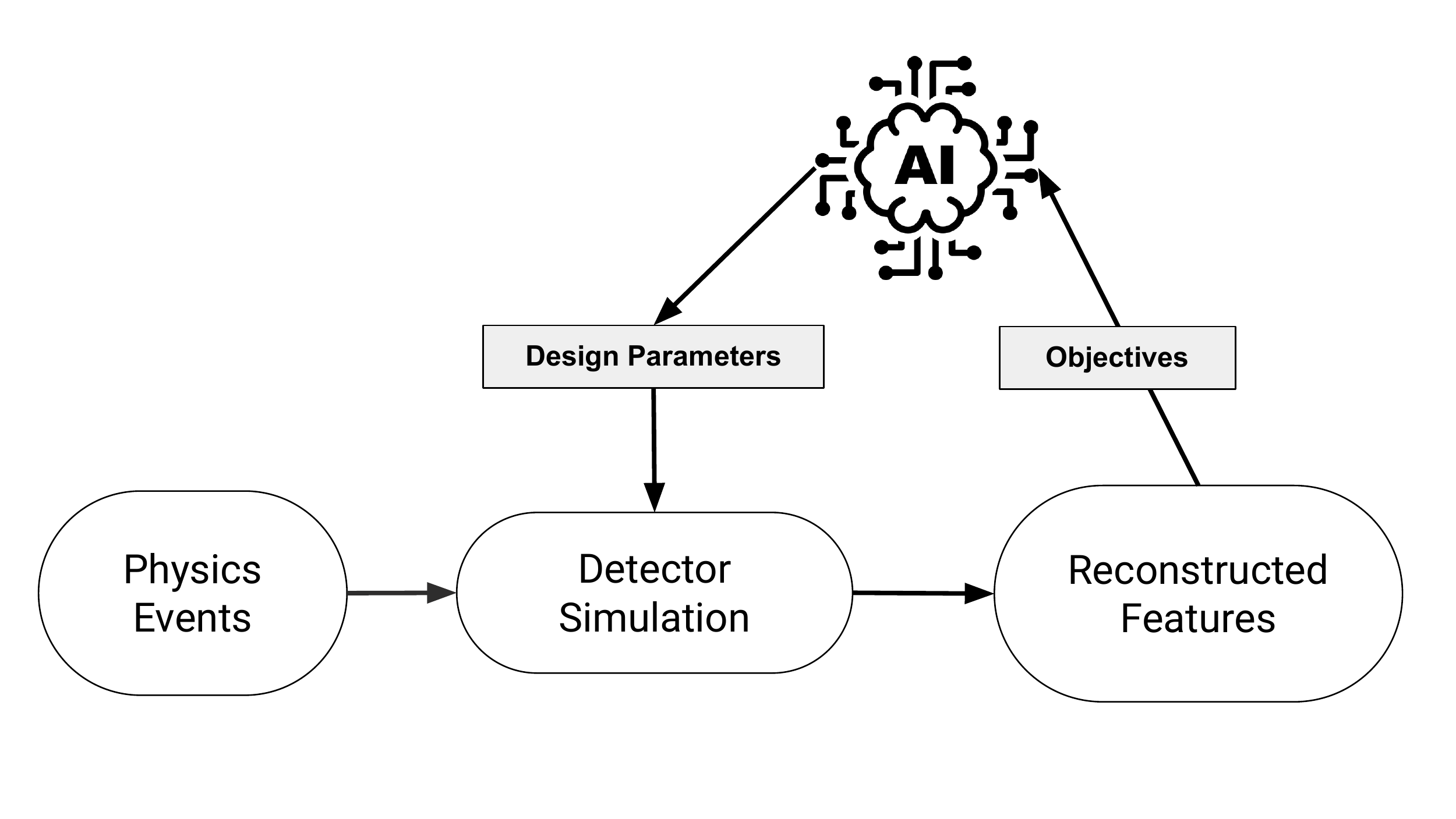}
    
    \caption{%(top) 
     \textbf{Workflow of detector design assisted by AI:} physics events are injected in a detector characterized by given design parameters. Reconstructed events are analyzed and figures of merit are quantified and passed to an AI-based strategy, which in turn suggests the next design point in this sequential approach; note that AI can also intervene in the simulation and reconstruction steps. 
    }
    \label{fig:design_AI}
\end{figure}

In this context, a workflow for detector design that has gained popularity in recent years \cite{cisbani2020ai} is represented by the schematic in Fig.~\ref{fig:design_AI}. 
It consists of a sequential AI-based strategy that collects information associated to previously generated design points, in the form of figures of merit (called \textit{objectives} in the following) that quantify the goodness of the design, and which suggests promising new design points for the next iteration. 

The ECCE AI Working Group achieved a continual multi-objective optimization (MOO) of the tracker design. 
Our approach deals with a complex optimization in a multidimensional design space (describing, \textit{e.g.}, geometry, mechanics, optics, etc) driven by multiple objectives that encode the detector performance, while satisfying several mechanical constraints. 
This framework has been developed in a way that can be easily extended to other sub-detectors or to a system of sub-detectors. 

\begin{figure}
    \centering
    \includegraphics[width=0.5\textwidth]{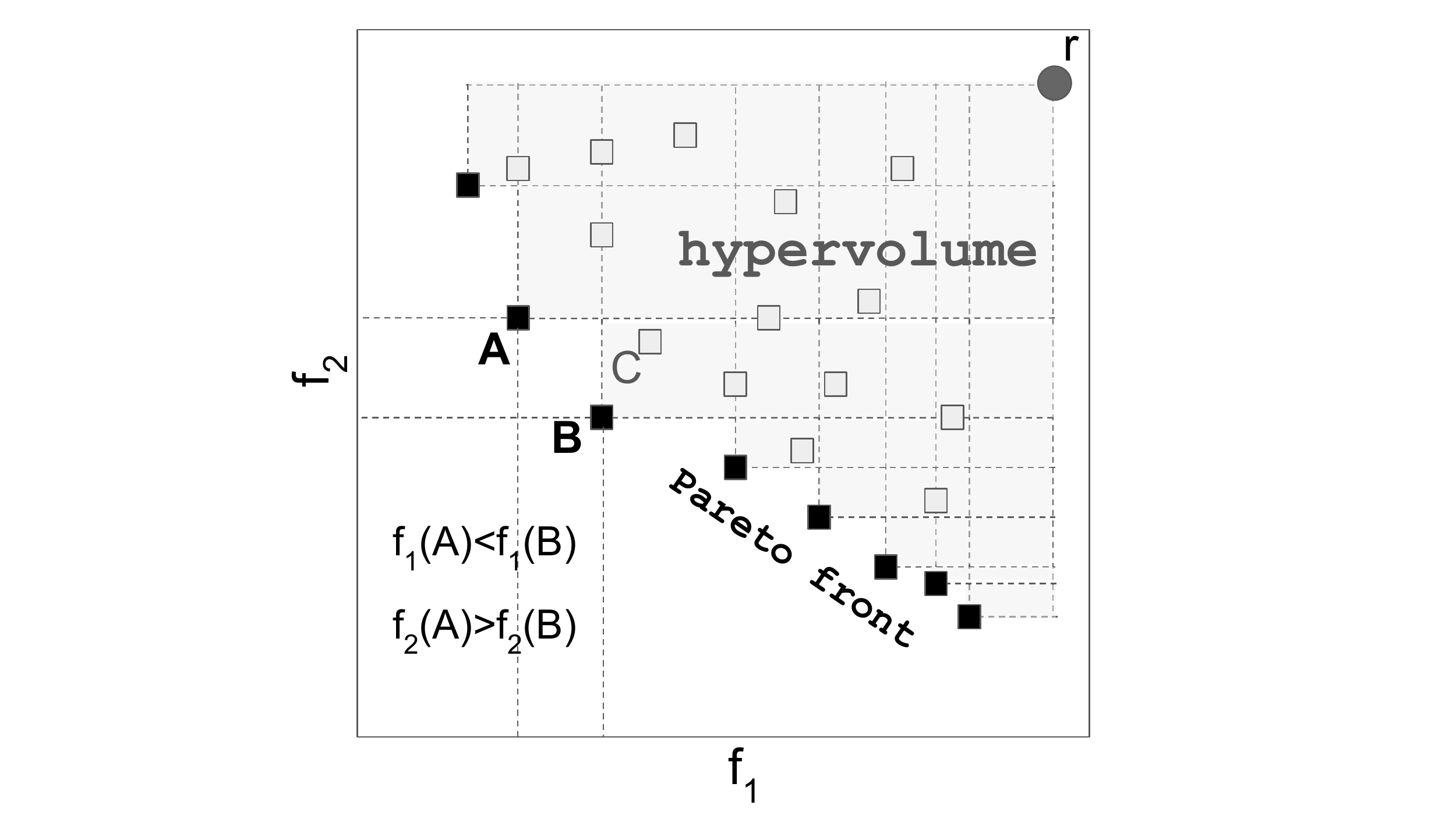}
    \caption{\textbf{Example of \textit{Pareto frontier} in a two-dimensional objective space:} The point C is not on the frontier and is dominated by both point A and point B. All the other points which are dominated by the Pareto frontier and that satisfy any constraints in the optimization problem are called \textit{feasible} solutions;
    The hypervolume is used as a metric for convergence, and is calculated with respect to a reference point $r$.
    }
    \label{fig:pareto_front}
\end{figure}

The definition of a generic MOO problem can be formulated as follows:
\begin{equation}\label{eq:moo}
\begin{aligned}
        min \ \mathbf{f_{m}}(\mathbf{x})  & \ \ \ \ \ \ m = 1, \cdots, M \\
       s.t. \ \ \ \mathbf{g_{j}}(\mathbf{x})\leq 0, & \ \ \ \ \ \ j = 1, \cdots, J \\ 
       \ \ \ \ \  \mathbf{h_{k}}(\mathbf{x})=0,     & \ \ \ \ \ \ k = 1, \cdots, K \\
       x_{i}^{L} \leq x_{i} \leq x_{i}^{U},         & \ \ \ \ \ \ i = 1, \cdots, N \\
\end{aligned}
\end{equation}
 where one has $M$ objective functions $f_{m}$ to optimize (\textit{e.g.}, detector resolution, efficiency, costs), subject to $J$ inequalities $g_{j}(x)$ and $K$ equality constraints $h_{k}(z)$ (\textit{e.g.}, mechanical constraints), in a design space of $N$ dimensions (\textit{e.g.}, geometry parameters that change the \geant design) with lower and upper bounds on each dimension.\footnote{Constraints are described later in Table \ref{tab:constraints}.} 
  Notice that overlaps in the design are checked before and during the optimization and are excluded by the constraints and ranges of the parameters.
 In solving these problems, one can come up with a set of \textit{non-dominated} or \textit{trade-off} solutions \cite{debreu1954valuation}, popularly known as Pareto-optimal solutions (see also Fig.~\ref{fig:pareto_front}). 

In this setting, we used a recently developed framework for MOO called pymoo \cite{blank2020pymoo} which supports evolutionary MOO algorithms such as Non-Dominated Sorting Genetic Algorithm (or NSGA-II, \cite{deb2002fast}).\footnote{The pymoo framework also supports other MOO approaches and a full list is documented in \cite{blank2020pymoo}.} The rationale behind this choice instead of, for example, principled approaches such as Bayesian Optimization \cite{cisbani2020ai}, emanates from the ECCE needs at the time of the detector proposal, such as the capability to quickly implement and run multiple parallel optimization pipelines implementing different technology choices and the possibility of dealing with non-differentiable objectives at the exploratory stage.  

\begin{figure}[!]
    \centering
    \includegraphics[width=0.45\textwidth, trim={0 2cm 0 0},clip]{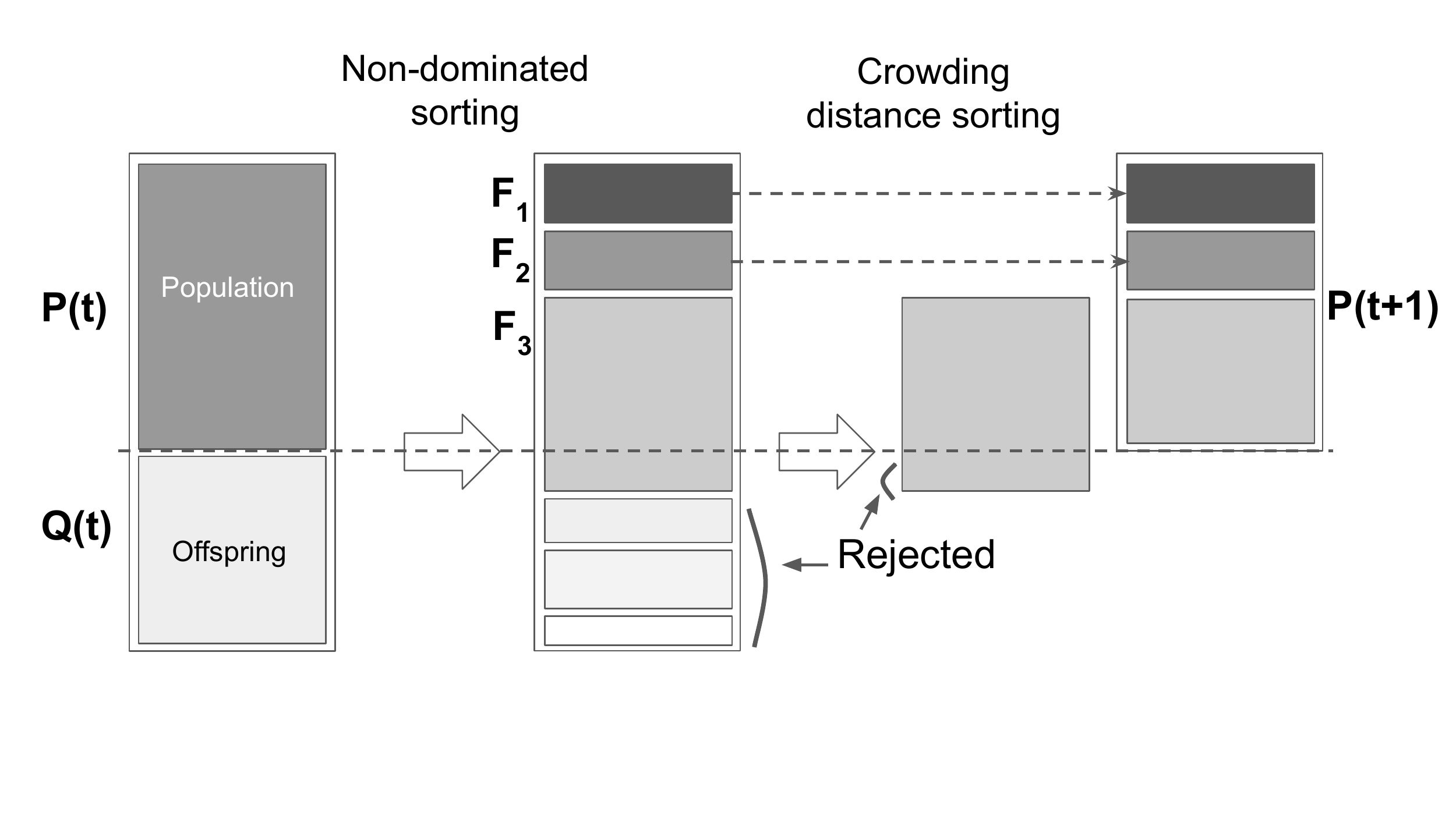}
    \caption{\textbf{The NSGA Workflow:} At time $t$, an offspring $Q(t)$ is created through a genetic algorithm~\cite{whitley1994genetic} from an $N-$sized population of design points $P(t)$. 
    The two populations are combined into an augmented population which is classified into different non-dominated classes $F_{i}$, starting from the first front $F_{1}$. 
    To restore the initial size of the population, the augmented space of solutions is trimmed. A metric called crowding distance is used to reject solutions and eventually provide an updated population of size $N$ at time $t+1$. 
    %Image taken from \cite{deb2002fast}.
    }
    \label{fig:NSGA_workflow}
\end{figure}

The NSGA workflow is described in Fig.~\ref{fig:NSGA_workflow}.
The main features of NSGA-II are (i) the usage of an elitist principle, (ii) an explicit diversity preserving mechanism, and (iii) ability of determining non-dominated solutions. 
The latter feature is of great importance for problems where objectives are of conflict with each other: that is an improved performance in an objective results in worse performance in another objective.
%
 %The pymoo framework also supports other MOO approaches and a full list is documented in \cite{blank2020pymoo}. 
 %
 For our purposes, we also tested NSGA-III which is suitable for the optimization of large number of objectives \cite{ishibuchi2016performance}.\footnote{For $\geq$ 4 objectives, NSGA-III is expected to perform better than NSGA-II.}

\begin{figure}[!]
    \centering
    \includegraphics[width=0.48\textwidth]{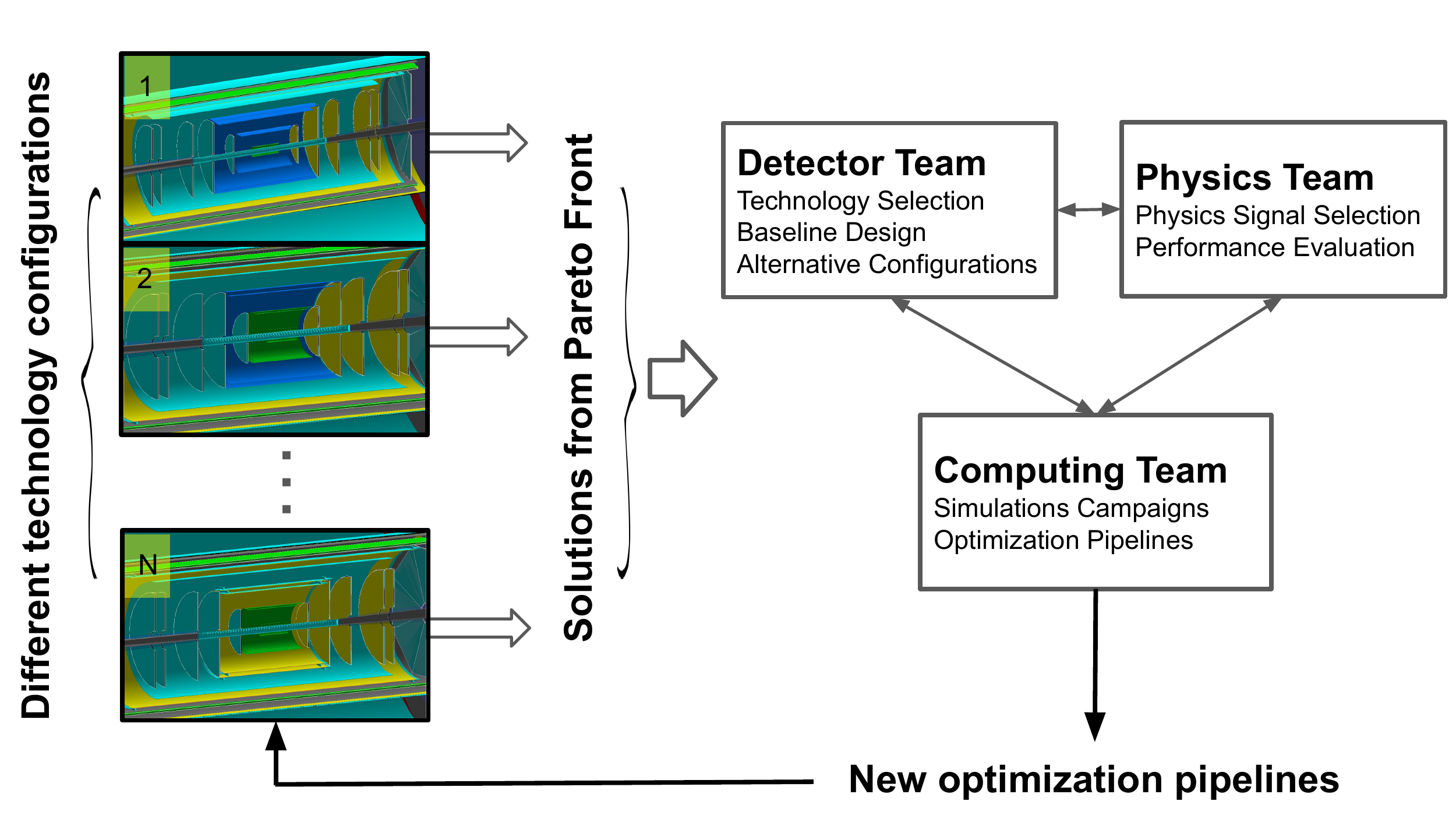}
    \caption{
    \textbf{Flowchart of continual optimization during proposal:} AI assists the design optimization process by providing insights and capturing hidden correlations among the design parameters. 
    This has been used during the entire detector proposal process to steer the design. At a given instant in time, N alternative candidate configurations are studied. For each we create an optimization pipeline, which results in a Pareto front of solutions. This new information helps steering the design: some configurations are rejected, while other ones (also dubbed `new references') are identified to potentially improve the design. New optimization pipelines are defined inspired by the new results and the process is iterated. During the design process, AI propelled the fundamental interplay between the ECCE Teams working on Physics, Detector and Computing. 
}
    \label{fig:pipeline_workflow}
\end{figure}

%As described in Sec.~\ref{sec:simulation}, the  simulation and reconstruction rely on the Fun4All framework \cite{fun4allGithub}. 
During the design optimization process of the tracking system, we used full \geant simulations of the entire ECCE detector. 
%In Sec.~\ref{sec:analysis} the reader can find a detailed description of the different parametrizations that have been utilized for several optimization pipelines of the inner tracker design optimization.
%
%Eventually the parametrization has been extended to include also the support structure in the design optimization problem.   
%Notice that overlaps in the design are checked before and during the optimization and are excluded by the constraints and ranges of the parameters.
%
%At least three objective functions characterizing the tracker performance (intrinsic detector response such as resolutions and efficiency) have been optimized simultaneously. The different designs have been optimized using particle gun samples of pions; after the optimization we used Pythia \cite{Sjostrand:2006za} samples to evaluate the improvement in the reconstruction of physics events (see Sec. \ref{subsec:physics_analysis}). 
%
AI played a crucial role in helping choose a combination of technologies for the inner tracker and was used as input to multiple iterations of the ECCE tracker design, which led to the current tracker layout. 
This was the result of a continual optimization process that evolved in time:
results were validated by looking at figures of merit that do not enter as objective functions in the optimization process (more details can be found in Sec. \ref{subsec:validation}); the decision making is left post hoc and discussed among the Computing, Detector and Physics teams.
%Fig.~\ref{fig:pipeline_workflow} displays a flowchart describing this continued optimization process that integrates AI and the interplay between the ECCE teams that took place during the detector proposal. 
A flowchart describing this continual optimization process is shown in Fig.~\ref{fig:pipeline_workflow}.

%
%AI provided insights in the design of the ECCE tracker by showing each time by how much the tracking system with different technology choices could be optimized. 
Ultimately this continual AI-assisted optimization led to a projective design after having extended the parametrized design to include the support structure of the inner tracker. The latter represents an ongoing R\&D project that is discussed in the next sections.

\begin{comment}
An example of these improvements is summarized in Fig. \ref{fig:improved_with_AI}.

\begin{figure}[!ht]
    \centering
    \includegraphics[width=0.45\textwidth]{figs/Proposal_AI_Improvement_2x2.pdf}
    \caption{The momentum dependence of the tracker momentum resolution for the ECCE reference tracker design (ECCE Simulation, blue solid circles) and for the projective mechanical support design of the ECCE ongoing project R\&D that will continue after the proposal (red solid circles). The latter shows a reduction of the impact of readout and services on the tracking resolution in the regions $1<\eta<1.5$ and $-1.5<\eta<1$. Note that the backward region relies on the EM calorimeter, and thus a resolution larger than the PWG requirement is acceptable. 
    \label{fig:improved_with_AI}
    }
\end{figure}
\end{comment}

 % Example of improvement of the ECCE tracker obtained with the AI-assisted design optimization strategy, compared to the 1$^{st}$ simulation campaign: (left) the momentum dependent momentum resolution in $|\eta|$ $<$ 0.5 for the costed ECCE Simulation and for the Projective design described in this section; (right) similarly, in 1.0 $<$ $|\eta|$ $<$ 1.5. The benchmark curve is taken from the Yellow Report and corresponds to an all silicon tracker with a 1.4 T magnetic field; more details can be found in~\cite{khalek2021science}.  

%%%%%%%%%%%%%%%%%%%%%%%%%%%%%%%%%%%%%%%%%%%%%%%%%
\section{ECCE Tracking System Simulation}\label{sec:simulation}
%%%%%%%%%%%%%%%%%%%%%%%%%%%%%%%%%%%%%%%%%%%%%%%%%

%\subsection{subsec:Software stack}

%\subsection{subsec:ECCE Central Detector }

\begin{figure*}[h!]
    \centering
    \includegraphics[trim=6em 0em 5em 0em, clip, width=0.48\textwidth]{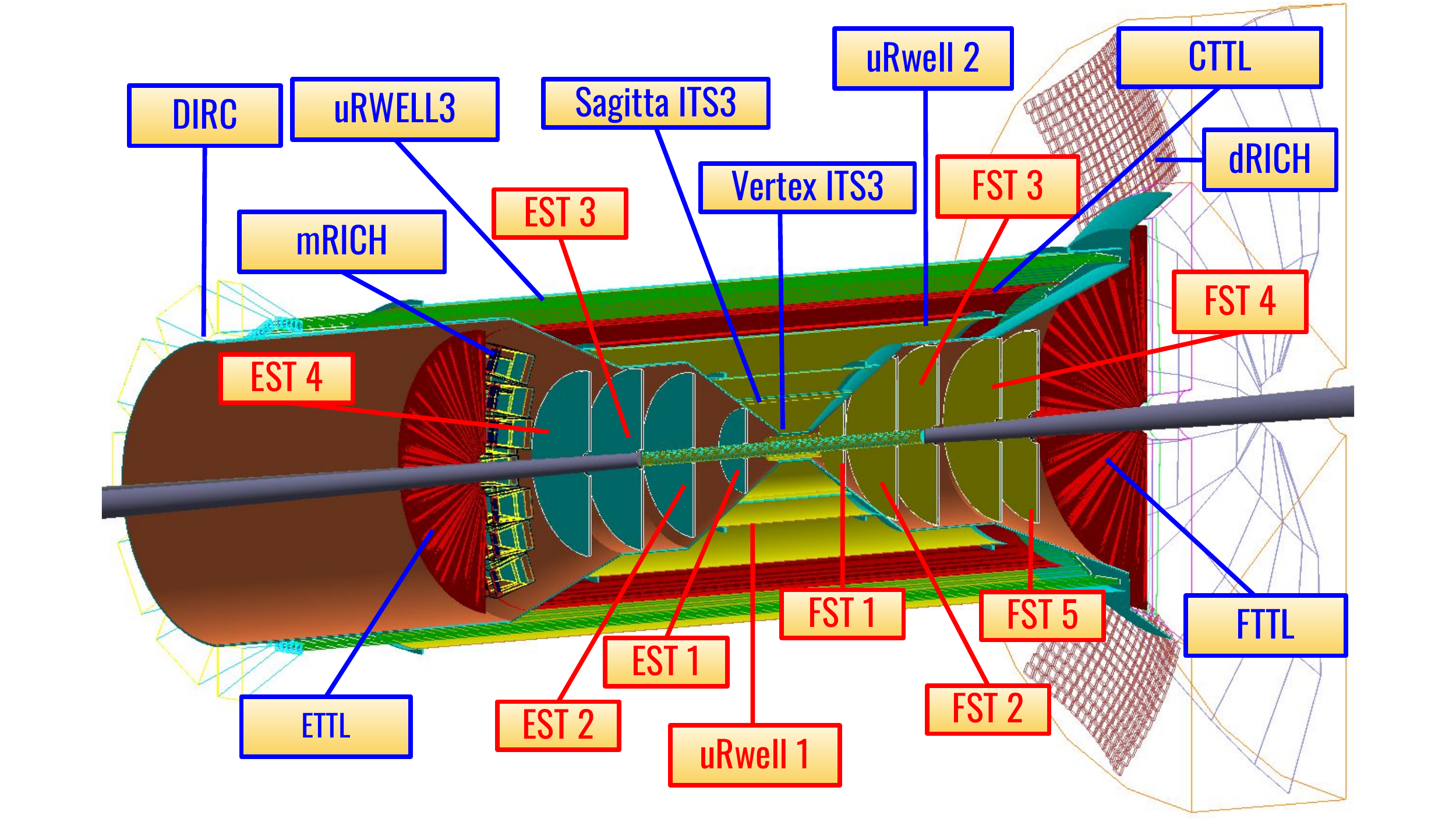}    
    \includegraphics[trim=6em 0em 5em 0em, clip, width=0.48\textwidth]{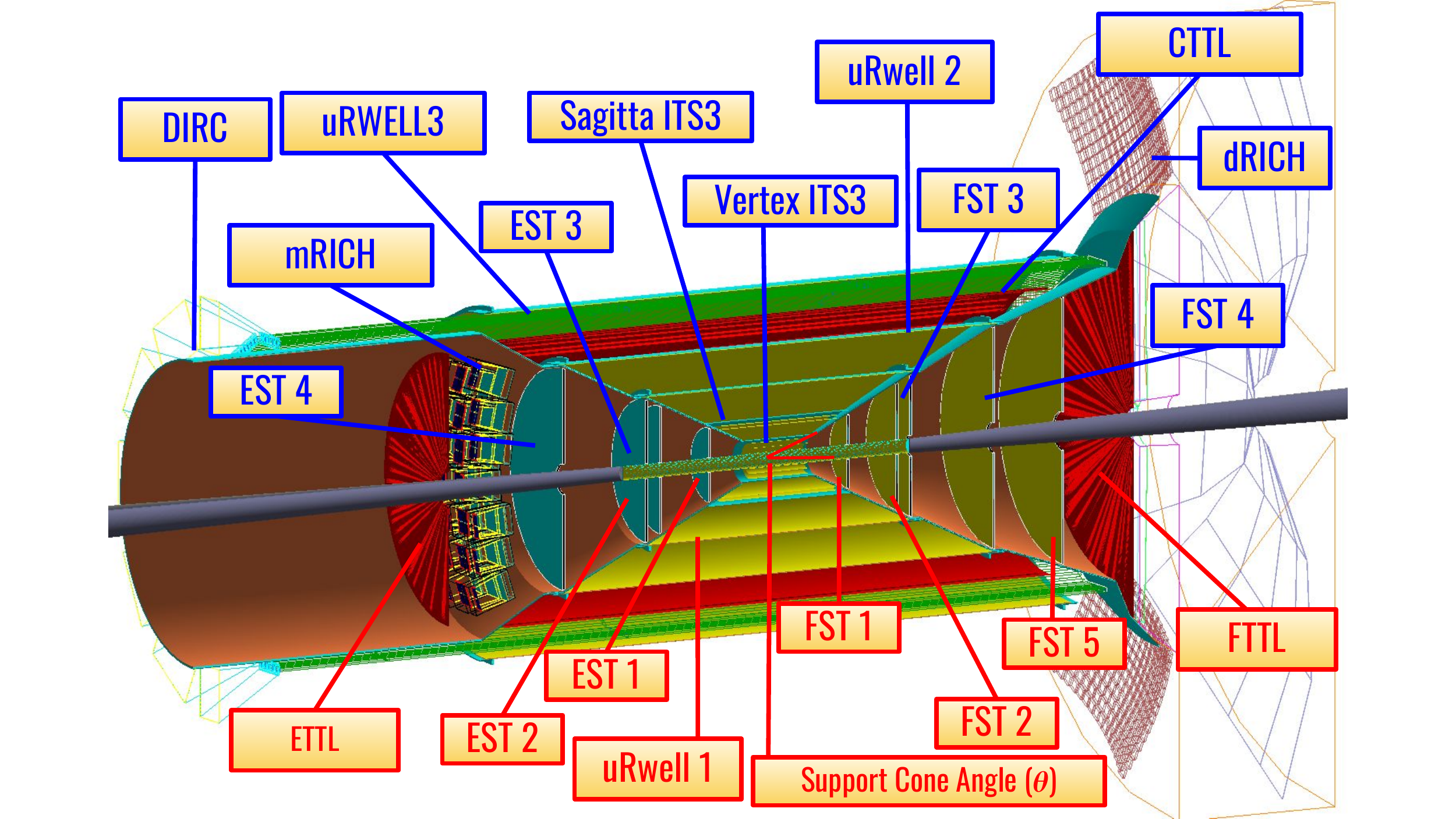}
 
    \caption{\textbf{Tracking and PID system in the non-projective (left) and the ongoing R\&D projective (right) designs:} the two figures show the different geometry and parametrization of the ECCE non-projective design (left) and of the ongoing R\&D projective design to optimize the support structure (right). 
    Labels in red indicate the sub-detector systems that were optimized, while the labels in blue are the sub-detector systems that were kept fixed due to geometrical constraint. The non-projective geometry (left) is a result of an optimization on the inner tracker layers (labeled in red) while keeping the support structure fixed, The angle made by the support structure to the IP is fixed at about $36.5\degree$. The projective geometry (right) is the result of an ongoing project R\&D to reduce the impact of readout and services on tracking resolution. 
    }
    \label{fig:ECCE_Tracking_non_proj_and_proj}
\end{figure*}

%“we created a Geant4~\cite{geant4} simulation based on the ECCE detector proposal~\cite{ecce-proposal}. The fun4all framework~\cite{f4a} was used to interface Geant4 with event generators”. And then authors could add another line saying what generator they used and any specific tunes.

The simulation and detector response shown in this document is based on \geant \cite{Agostinelli:2002hh} and was carried out using the Fun4All framework \cite{fun4allGithub, ecce-note-comp-2021-01}. 
%The ECCE-EIC core software utilizes daily, weekly builds and tagged production releases \cite{ecce_eic_sw}. 
%%
%%Development of the framework was driven by reconstruction and analysis needs where the incoming data had to be processed quickly. The framework is extremely modular such that simulations of various detector subsystems can be independently developed and can be integrated within the framework with ease.

%%Fig.~\ref{fig:ECCE_Workflow} represents a schematic of the ECCE Software. 

%Various repositories hold relevant Fun4All analysis code. The macros combine the needed packages into one root script to run jobs. 
%Default macros can run a wide variety of event generators, detector configurations, etc.
%One can pick and choose what to run depending on the analysis needs. The framework also allows the creation of the relevant module and registers it with Fun4All. 

%A wide range of event generators (\textit{e.g.}, HIJING \cite{wang1991hijing}, PYTHIA \cite{sjostrand2008brief}, Sartre \cite{toll2014dipole}, single particle, etc) is supported through the EIC-smear interface and Fun4All. 
The optimization pipelines are based on particle gun samples of pions, where we used $\pi^{-}$ and tested that the performance with $\pi^{+}$ were consistent. Performance in the electron-going direction was also checked post-hoc with particle gun samples of electrons. 
The improved performance is further validated with physics analyses,  using the datasets generated during the ECCE simulation campaigns; in Sec.~\ref{sec:analysis} we show in particular results based on semi-inclusive deep inelastic scattering (SIDIS) events.

The ECCE tracking detector \cite{ecce-paper-det-2022-03}, represented in Fig. \ref{fig:ECCE_Tracking_non_proj_and_proj} (left), consists of different layers in the barrel and the two end-caps, and is tightly integrated with the PID detectors:

(i) The silicon vertex/tracking detector is an ALICE ITS-3 type high precision cylindrical/disk vertex tracker \cite{ALICEITSproject:2021kcd, colella2021alice}) based on the new Monolithic Active Pixel Sensor (MAPS); the barrel detector consists of 5 MAPS layers; the silicon hadron endcap consists of 5 MAPS disks; and the silicon electron endcap has 4 MAPS disks.

(ii) A gas tracking system is based on $\mu$Rwell technology, that is a single-stage amplification Micro Pattern Gaseous Detector (MPGD) that is a derivative of the Gas Electron Multiplier (GEM) technology. In ECCE $\mu$Rwell layers will form three barrel tracking layers further out from the beam-pipe than the silicon layers; namely, two inner-barrel layers and a single  outer-barrel $\mu$Rwell  layer.   All $\mu$Rwell detectors will have 2D strip based readout. The strip pitch for all three layers will be 400 $\mu$m.

(iii) The tracking system is completed by AC-LGAD-based time of flight (TOF) detectors providing additional  hit information for  track reconstruction as well. In the central region a TOF (dubbed CTTL) is placed behind the high-performance DIRC (hpDIRC); in the hadron-going side a TOF (dubbed FTTL) is placed before the dual RICH (dRICH) and a $\mu$Rwell placed after the dRICH; in the electron-going direction a $\mu$Rwell layer is placed before the modular RICH (mRICH), which is followed by a TOF later (dubbed ETTL).

%The simulation of the ECCE tracking system is shown in Fig.~\ref{fig:ECCE_Tracking_non_proj_and_proj}. %
An important consideration for all large-scale detectors is the provision of readout (power and signal cables) and other services (e.g., cooling). Clearly the aim is to minimize the impact of readout and services in terms of affecting the detector's acceptance or tracking resolution, for example.  This effort is ongoing R\&D for the project.

In the following sections, the reader can find more details on the implementation of the optimization pipelines and utilized computing resources.

\section{Analysis Workflow}\label{sec:analysis}
%%%%%%%%%%%%%%%%%%%%%%%%%%%%%%%%%%%%%%%%%%%%%%%%%

The optimization of the ECCE-tracking system \cite{ecce-paper-det-2022-01, ecce-paper-det-2022-03} has been characterized by two main phases during which the sub-detectors composing the tracker evolved into more advanced renditions.

\paragraph{Phase I optimization}\footnote{Phase I corresponds to a timeline between June-2021 to Sept-2021. Preliminary studies done between March-2021 to May-2021 are not reported here.} The \geant implementation of the detectors were at first simplified, \textit{e.g.}, detector modules were mounted on a simplified conical support structure made of aluminum. 
%The support structure was initially not implemented in the optimization. 
The optimization pipelines consisted of symmetric arrangement of detectors in the electron-going and hadron-going directions (5 disks on each side). The DIRC detector for PID in the barrel region was modelled with a simple geometry made by a cylinder and conical mirrors.
AC-LGAD-based TOF detectors were modelled as simplified silicon disks at first; %more detailed simulations of the outer trackers have been developed during this phase.
the outer trackers had more fine-grained simulations implemented, with realistic support structures and services implemented.
The optimization pipelines included various combinations of detector technologies for the inner trackers. At the end of this phase, a decision on the choice of the barrel technology and the disk technologies was made using the AI results. 

%\cris{At the end of this section show the two figures of the non-projective and R\&D projective obtained with this work.}

%\cris{
%The ECCE tracking and PID system during Phase-I is shown in Fig.~\ref{fig:Phase-I_ECCEDetector}, which also displays the non-projective geometry of the support structure. 
%}

\begin{comment}
DIRC bars were simulated. Before that it was a simple cylinder and G4Tub
CTTL, ETTL, FTTL
I am not so sure of the details here but this was fine grained simulations (a new class called as g4ttl/PHG4TTLSubsystem was introduced) ,  But before (phase-1) it was just silicon layers and resembled to that of FSTs (but with 30micron resolution)
uRwell
Mainly had support rings
\end{comment}

\paragraph{Phase II optimization}\footnote{Phase II corresponds to optimization pipelines that run from Sept-2021 to Nov-2021.} 
These pipelines had a more realistic implementation of the support structure incorporating cabling, support carbon fiber, cooling system, etc. 
More detailed simulation of the PID Detectors (\textit{e.g.}, DIRC bars and dRICH sub-systems) were integrated as well as fine-grained simulations of TTL layers (CTTL, ETTL, FTTL) previously simulated as simple silicon layers modules. %
More stringent engineering constraints were considered such as the sensor size for MAPS detector (ITS3). This phase also considered an asymmetric arrangement of the detectors in the endcap regions, with a maximum of 4 EST disks in the electron-going end-cap and 5 FST disks in the hadron-going endcap: 
%The placement of detectors in the endcap regions are asymmetric for the pipelines that were done during Phase-II. 
due to this asymmetric spatial arrangement, the angle subtended by detectors in the two endcap regions could be varied. %An optimization was carried out with asymmetric spatial arrangement of EST and FST disks. 
This eventually developed into the idea of a projective geometry in a pipeline that characterizes an ongoing R\&D project for optimizing the design of the support structure.

A detailed description of the most recent parametrization used for the detector proposal can be found in \ref{app:par_tracking}, 
%Such parametrization has been used for the optimization of the non-projective design, which had the inner tracker support structure fixed. 
 along with the parametrization used in an ongoing R\&D project to optimize the support structure of the inner tracker. %;
 %has been modified to explore the possibility of including in the optimization the design of the support structure. This resulted in the R\&D project always described in \ref{app:par_tracking}; 
% a schematic representation is displayed in Fig.~\ref{fig:SupportStructureparametrization}. 

Fig.~\ref{fig:ECCE_Tracking_non_proj_and_proj} shows a comparison of the ECCE reference non-projective design and the projective design from the ongoing R\&D, both of which resulted from the AI-assisted procedure described in this paper.

\subsection{Encoding of Design Criteria}\label{subsec:encoding}
%%%%%%%%%%%%%%%%%%%%%%%%%%%%%%%%%%%%%%%%%%%%%%%%%

%\cris{Talk about only phase-II optimization; phase I mention only it had a symmetric parametrization and perhaps point to the analysis note for details. This reduces also the tables later.}

Design criteria need to be encoded to steer the design during the optimization process. 
For each design point we need to compute the corresponding objectives $f_m$, namely the momentum resolution, angular resolution, and Kalman filter efficiency.

We will refer in the following only to the more recent Phase II optimization.\footnote{Similar considerations apply also for Phase I optimization.} 
Phase II has been characterized by two types of optimization pipelines: the first used a parametrization of the inner tracker during the optimization process and led to the ECCE tracker non-projective design; the second branched off the first as an independent R\&D effort that included the parametrization of the support structure and led to a projective design.

Details on the two types of optimization pipelines can be found in the following tables: 
Table~\ref{tab:hyperparameters} describes the main hyperparameters and the dimensionality of the optimization problem, in particular of the design space and the objective space;  Table~\ref{tab:ranges} reports the range of each design parameter\footnote{The design points are normalized in the range [0-1], using a min-max scaler $x_{i} = x (x_{max} - x_{min}) + x_{min}$, where $x_i$ is the normalized design point with a un-normalized design point $x$ generated between the range $[x_{min}, x_{max}]$.};
Table~\ref{tab:constraints} summarizes the constraints for both the non-projective and projective geometries. 
We also considered in our design a safe minimum distance between the disks of 10 cm and include a constraint on the difference between the outer and inner radii of each disk, namely R$_{out}$ - R$_{in}$, to be a multiple of the sensor cell size (17.8 mm $\times$ 30.0 mm), see Table~\ref{tab:constraints}.
These constraints are common to the non-projective and the projective designs.
For more details on the parametrizations and on the corresponding detector performance the reader can refer to \ref{app:par_tracking} and  \ref{app:comparison}, respectively. 
%In what follows we provide a thorough description of the design criteria that have been encoded in the optimization strategy. 

%%%%%%%%%%%%%%%%%%%%%%%%%%%%%%%%%%%%%%%%%%%%%%%%%%%%%%%%%%%%

\begin{table}[!bh!]
   
    \centering
    \begin{tabular}{|M{3.cm} | M{1.5cm}| M{2.0cm}|}
        \hline
        \textbf{description} & \textbf{symbol} & \textbf{value}\\ 
        \hline
        \hline
        \small{population size} & N & 100 \\
        \small{\# objectives}   & M & 3\\ 
        \small{offspring} & O & 30\\
        \small{design size} & D & 11 (9) \\
        \small{\# calls (tot. budget)} & $-$ & 200\\
        \small{\# cores} & $-$ & \small{same as offspring} \\
        \small{\# charged $\pi$ tracks} & N$_{\textup{trk}}$ & 120k\\
        \small{\# bins in $\eta$} & N$_{\eta}$ & 5\\
        \small{\# bins in p} & N$_{\textup{p}}$ & 10\\
        \hline
    \end{tabular}
    \caption{\textbf{Summary of the hyperparameters of the design optimization:}
    the values reported in the table have been used during the optimization of the non-projective design of the ECCE tracker. For completeness and when they differ from the non-projective case, we also report in parentheses the values corresponding to the ongoing R\&D project for the projective design of the support structure. 
    \label{tab:hyperparameters}
    }
    
\end{table}

\begin{table}[!]
    \centering
    %\resizebox{\columnwidth}{!}{
    \begin{adjustbox}{width=0.9\columnwidth}
    %\begin{tabular}{| c | c | c |}
    \begin{tabular}{|M{5.cm}|M{3.5cm}|}
        \hline 
        \rowcolor{blue!15}\multicolumn{2}{|c|}{  \textbf{ECCE design (non-projective)}}\\
        \hline
        \textbf{Design Parameter} & \textbf{Range} \\
        \hline
        $\mu$RWELL 1 (Inner) ($r$) Radius  & [17.0, 51.0 cm] \\
        \hline
        $\mu$RWELL 2 (Inner) ($r$) Radius  & [18.0, 51.0 cm] \\
        \hline
        EST 4 $z$ position &   [-110.0, -50.0 cm] \\
        \hline
        EST 3 $z$ position &   [-110.0, -40.0 cm] \\
        \hline
        EST 2 $z$ position &   [-80.0, -30.0 cm] \\
        \hline
        EST 1 $z$ position &   [-50.0, -20.0 cm] \\
        \hline
        FST 1 $z$ position &  [20.0, 50.0 cm] \\
        \hline 
        FST 2 $z$ position &  [30.0, 80.0 cm] \\
        \hline
        FST 3 $z$ position &  [40.0, 110.0 cm] \\
        \hline
        FST 4 $z$ position &  [50.0, 125.0 cm] \\
        \hline
        FST 5 $z$ position &  [60.0, 125.0 cm] \\
        \hline
        \rowcolor{blue!15}\multicolumn{2}{|c|}{  \textbf{ECCE ongoing R\&D (projective)}}\\
        \hline
        \textbf{Design Parameter}  & \textbf{Range} \\
        \hline
        Angle (Support Cone)   & [25.0$\degree$, 30.0$\degree$] \\
        \hline
        $\mu$RWELL 1 (Inner) Radius  & [25.0, 45.0 cm] \\
        \hline
        ETTL $z$ position  & [-171.0, -161.0 cm] \\
        \hline
        EST 2 $z$ position  & [45, 100 cm] \\
        \hline
        EST 1 $z$ position  & [35, 50 cm] \\
        \hline
        FST 1 $z$ position  & [35, 50 cm] \\
        \hline 
        FST 2 $z$ position  & [45, 100 cm] \\
        \hline
        FST 5 $z$ position  & [100, 150 cm] \\
        \hline
        FTTL $z$ postion  & [156, 183 cm] \\
        \hline
        
    \end{tabular}
    \end{adjustbox}
    \caption{\textbf{Ranges of the design parameters:}  the table summarises the design points that are optimized for the non-projective and the projective case. The optimization range is also given for each parameter. The parameters corresponding to the non-projective case were optimized sequentially over at least three iterations, with each iteration having a set of parameters fixed. For instance, the EST/FST disks were optimized during the first iteration assuming symmetric design, and for the subsequent iterations the EST disks were optimized further in the electron endcap region.  
    %The other geometric parameter for each of the detector systems are calculated based on the parametrization discussed above. 
    Along with the design parameters the design comes with geometrical constraints too. 
    These are implemented as strong and soft constraints in the Table \ref{tab:constraints}. 
    \label{tab:ranges}
    }
\end{table}

\begin{table}[!]
    \centering
    \resizebox{\columnwidth}{!}{%
    \begin{tabular}{|M{3.0cm}|M{5.7cm}|M{4.8cm}|}
        \hline
         \textbf{sub-detector} & \textbf{constraint} & \textbf{description} \\
        \hline
        \hline
        EST/FST disks  & %$\sum_{i} \frac{(RMax_{i} - RMin_{i})}{L} = \mathbb{N}$ 
        
        $$min \Bigg\{ \sum_{i}^{disks} \Bigg| \frac{R^{i}_{out} - R^{i}_{in}}{d} - \floor*{\frac{R^{i}_{out} - R^{i}_{in}}{d}} \Bigg| \Bigg\}$$
        
        & \textbf{soft constraint}: sum of residuals in sensor coverage for disks; sensor dimensions: $d$ = 17.8 (30.0) mm\\
        \hline 
        
        %FST/EST Disks & $\sum_{i} \frac{(RMax_{i} - RMin_{i})}{1.8} = \mathbb{N}$ & MAPS sensor width is 1.78 cm. Soft Constraint\\
        \hline
        EST/FST disks & $z_{n+1} - z_{n} >= 10.0$ cm & \textbf{strong constraint}: minimum distance between 2 consecutive disks   \\
        \hline
        sagitta layers & $$ min\Bigg\{ \Bigg| \frac{2 \pi r_{sagitta}}{w} - \floor*{\frac{2 \pi r_{sagitta}}{w}} \Bigg| \Bigg\}$$ &  \textbf{soft constraint}: residual in sensor coverage for every layer; sensor strip width: $w$ = 17.8 mm\\
        \hline
        $\mu$RWELL & $r_{n+1} - r_{n} >= 5.0$ cm &  \textbf{strong constraint}: minimum distance between $\mu$Rwell barrel layers\\
        \hline
        
    \end{tabular}
    }
    \caption{\textbf{Constraints in the design optimization:} the table summarises constraints for both the projective and the non-projective designs. 
    %The constraints are from the FST/EST disks since they are made by tiling up MAPS pixels. 
    %The pixels will have fixed dimensions. 
    Soft constraints are constraints that can be violated to a certain degree by the MOO and then one can quantify the degree of unfeasibility (see \cite{blank2020pymoo}). Also, minor adjustments on tiling up the pixels can be done post hoc optimization such that the detector geometry is realisable. Strong constraints during the optimization designs rest in high penalties since these constraints cannot be violated. %\textit{e.g.}, the distance between the disks cannot be closer than 10 cm for servicing requirement.
    %The constraints in this table are common to all designs.
    Two additional constraints are implicitly taken into account by internal parametrization and reflected in the \geant design: the inner radii of the disks in the endcap and the inner vertex layer in the barrel have to be compatible with the beam envelope dimensions; the barrel layers lengths and the outer radii of the disks have to be compatible with an tracking support structure \cite{arrington2021eic}. %These are provided by internal parametrization and reflected in the Geant4 design.
    Potential overlaps among modules are checked before and during the optimization.
    }
    \label{tab:constraints}
\end{table}

%%%%%%%%%%%%%%%%%%%%%%%%%%%%%%%%%%%%%%%%%%%%%%%%%%%%%%%%%%%%

The objectives depend on the kinematics and are calculated in 5 main bins in pseudorapidity ($\eta$):  (i) -3.5 $ \leq \eta < $ -2.0 (corresponding to the  electron-going direction), (ii) -2.0 $ \leq \eta <$ -1.0 (the transition region in electron-going direction), (iii) -1 $ \leq \eta <$ 1 (the central barrel),  (iv) 1 $ \leq \eta <$ 2.0 (the transition region in the hadron-going direction) and (v) 2.0 $ \leq \eta <$ 3.5 (the hadron-going direction). 
The rationale behind this binning is a combination of different aspects: the correspondence with the binning in the EIC Yellow Report \cite{khalek2021science}, the asymmetric arrangement of detectors in electron-going and hadron-going directions and the division in pseudorapidity between the barrel region and the endcap. %\footnote{A finer binning is of course possible but it comes at an increase of computing time needed to run the simulations.} 
Particular attention is given to the transition region between barrel and endcaps as well as at large $|\eta| \sim$ 3.5 close to the beamline. 

Charged pions are generated uniformly in the phase-space that covers the range in momentum magnitude $p$~$\in$~[0,20]~GeV/c and the range in pseudorapidity $\eta$~$\in$~(-3.5,3.5). 
Each bin in $\eta$ is further subdivided in 15 bins in momentum $p$. 
%
%Studies have been performed to determine the number of tracks (N$_{trk}$) that we need to simulate in order to get sufficiently small statistical uncertainty
 For each design point we simulated N$_{trk}=$120k charged pions.\footnote{From Phase I to Phase II, the design became asymmetric in the two endcaps, therefore we needed to extend the $\eta$-coverage and increase the statistics. The momentum range was reduced to [0,20] GeV/c to optimize the computing budget.} 
This number ensured large enough statistics over the entire phase space and the stability of the fits in all of the bins of Eqs.~\eqref{eq:objectives_reso}.

In order to calculate the relative momentum (cf. Fig.~\ref{fig:Phase-II_Momentum_res}) and absolute angular resolution (cf. Fig~\ref{fig:Phase-II_dth_res}) we fit the following objectives: 

\begin{equation}\label{eq:obj_definition}
\begin{split}
    \frac{\Delta p}{p} &= \frac{p_{reco}-p_{true}}{p_{true}} \\ 
    \Delta \theta  &= \theta_{reco}-\theta_{true} 
\end{split}
\end{equation}

\begin{comment}
To calculate the angular resolution at the PID location (cf. Fig.~\ref{}) we do fit the following objectives:

\begin{equation}\label{eq:obj_definition_PID}
\begin{split}
    d\phi_{proj} &= \tan^{-1}{\Big(\frac{(\Vec{p}_{reco} - \Vec{p}_{true}) \cdot (\Vec{p}_{true}\times \Vec{z}) \times \Vec{p}_{true}}{|\Vec{p}_{true}| |(\Vec{p}_{true}\times \Vec{z}) \times \Vec{p}_{true}|}\Big)} \\ 
   d\theta_{proj} &= \tan^{-1}{\Big(\frac{(\Vec{p}_{reco} - \Vec{p}_{true}) \cdot (\Vec{p}_{true}\times \Vec{z})}{|\Vec{p}_{true}| |(\Vec{p}_{true}\times \Vec{z})|}\Big)} \\
   \Vec{p}_{true} &= True~\Vec{p}~at~PID \\
   \Vec{p}_{reco} &= Reconstructed~\Vec{p}~at~PID
\end{split}
\end{equation}
\end{comment}

\begin{figure}[!]
    \centering
    \includegraphics[width = 0.4\textwidth]{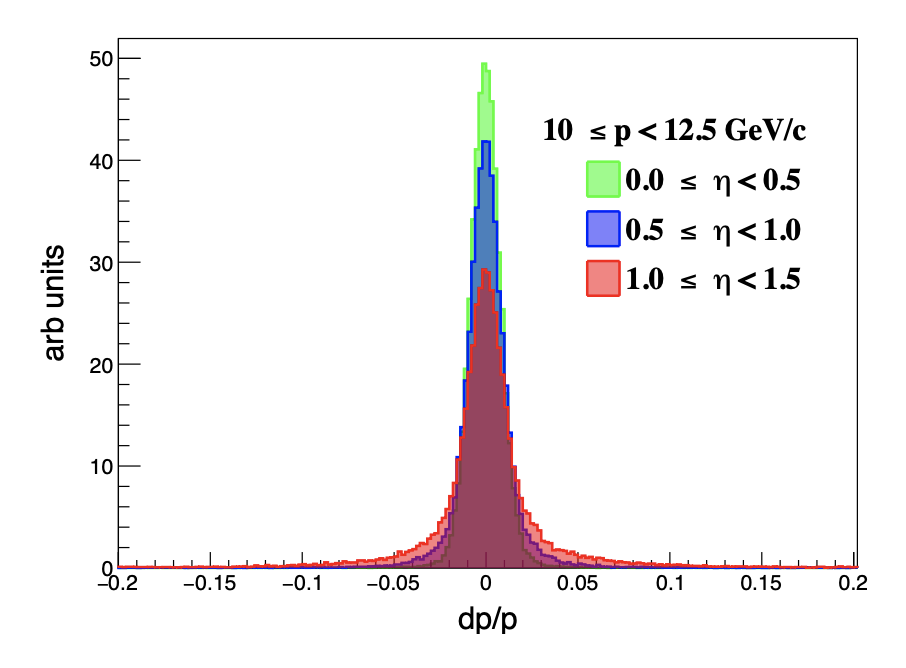}
    \caption{\textbf{Fit strategy: } a double-Gaussian fit function is utilized to extract the resolutions. Such a fit function provided good reduced $\chi^{2}$ and more stable extractions compared to single-Gaussian fits. The resolution is obtained as an average of the two $\sigma$'s weighted by the relative areas of the two Gaussians according to Eq.~\eqref{eq:DG}. The figure represents the results corresponding to a particular bin in $\eta$ and $p$. 
    }
    \label{fig:double_gauss}
\end{figure}

Following the definitions of Eq.~\eqref{eq:obj_definition}, histograms of the relative momentum resolution and the absolute angular resolution are produced for each bin in $\eta$ and $p$ and the corresponding fits are calculated. % as shown in Fig.~\ref{fig:fit_results}.
Using single-Gaussian (SG) fits (also utilized in the Yellow Report \cite{khalek2021science}) implies systematically better resolutions but worse reduced $\chi^{2}$: therefore we decided to utilize double-Gaussian (DG) fits, as shown in Fig.~\ref{fig:double_gauss}. 
This provided a more robust fit strategy. 
The reduced $\chi^{2}$ range with DG fits ranges from 1.2 to 2.8 at most, with the majority of the fits stable at lower values. 
The largest numbers correspond either to the transition between the barrel and endcaps---where tracks cross more material in the non-projective design---or to large pseudorapidity, particularly close to the inner radii of the disks. By using SG fits, the reduced $\chi^{2}$ values can be as large as 10-20 in the transition region. A detailed study comparing SG to DG fits is shown in Fig.~\ref{fig:Phase-II_Momentum_res}.

The final DG resolution has been defined as an average of the two $\sigma$'s weighted by the relative areas of the two Gaussians:\footnote{A different definition could be based on the weighted average of the variances $\sigma_{1,2}^{2}$ to obtain the final variance $\sigma_{DG}^{2}$. This typically implied a few \% relative difference on the final value of $\sigma_{DG}$ which has been considered a negligible effect.}
\begin{equation}\label{eq:DG}
    \sigma_{DG} \sim \frac{A_{1}\sigma_{1}+A_{2}\sigma_{2}}{A_{1}+A_{2}}
\end{equation}

The results obtained for the resolutions in each bin corresponding to each new design point are divided by the values corresponding to the baseline design, so that in each bin a ratio $R(f)_{p,\eta}$ is provided. 
Finally a weighted sum of these ratios is performed to build a global figure of merit (for both the relative momentum and the angular resolutions):  
\begin{equation}\label{eq:objectives_reso}
    R(f) = \frac{1}{N_{\eta}}\sum_{\eta} \left( \frac{\sum_{p} w_{p,\eta} \cdot R(f)_{p,\eta}}{\sum_{p} w_{p,\eta}} \right)
\end{equation}
where the objective function $f$ is either the momentum or the angular resolution described by Eq.~\eqref{eq:obj_definition}, and the weight $w_{p,\eta} = 1/\sigma^{2}(f_{p,\eta})$ is calculated in each $\eta, p$ bin and it is proportional to the inverse of the variance corresponding to the objective functions $f$.
%%either Eq.~\eqref{eq:obj_definition_PID} or

An additional objective function has been included in the optimization problem: this is a global objective function corresponding to the fraction of tracks that are not reconstructed by the Kalman filter (KF \cite{FRUHWIRTH1987444}), or equivalently the KF inefficiency:
\begin{equation}\label{eq:objective_KF}
    R(\textup{KF}) = \frac{\textup{N(tracks \ not \ reconstructed)}}{\textup{N(total number of tracks)}}
\end{equation}

Following the above definitions, the design problem corresponds to minimizing the objective functions defined in Eqs.~\eqref{eq:objectives_reso}, \eqref{eq:objective_KF}: a weighted average ratio smaller than one and a KF inefficiency as close as possible to zero correspond to an improvement in performance compared to the baseline design.
%; as for the KF inefficiency, we ideally want to find a design point that pushes this value as close as possible to zero. 

%\vspace{5cm}

%\clearpage
%%%%%%%%%%%%%%%%%%%%%%%%%%%%%%%%%%%%%
%\subsection{Results}
\subsection{Convergence and Performance at Pareto Front}
%%%%%%%%%%%%%%%%%%%%%%%%%%%%%%%%%%%%%

\begin{figure}[!]
    \centering
    \includegraphics[width=0.48\textwidth]{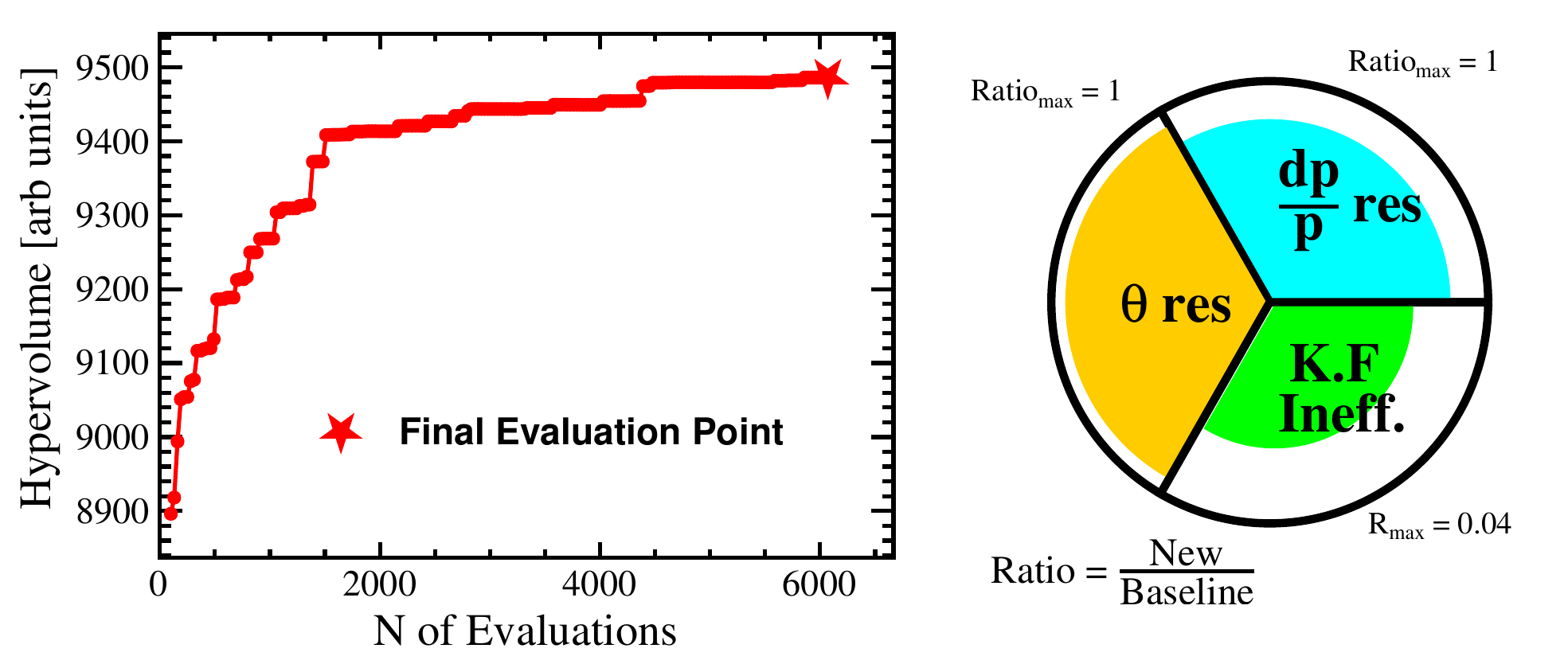}
    \caption{\textbf{Metric for convergence and performance at Pareto front:} (left) The hypervolume can be used as a metric for convergence. Checkpoints are created during the optimization and snapshots of the evolving designs are saved. (right) A petal diagram with the three objectives corresponding to one solution in the Pareto front. The momentum and angular resolutions are expressed as ratios with respect to a baseline design to improve; the KF inefficiency is taken as an absolute value. An optimal design optimizes all of the above defined objectives. 
}
    \label{fig:convergence}
\end{figure}

%-------------------------------------------
%\paragraph{Convergence and Pareto solutions:}
%
 We remind the reader that the Pareto front is the set of trade-off solutions of our problem. Fig.~\ref{fig:convergence} shows the convergence plot obtained utilizing the hypervolume as metric in the objective space.\footnote{Early stopping can occur if no change in the hypervolume is observed after a certain number of evaluations.} A petal diagram is used to visualize the values of three objectives corresponding to one of the solutions extracted from the Pareto front. 
 
Checkpoints are created to store the NSGA-II-updated population of design points. A survey of the detector performance is created after each call to monitor potential anomaly behavior of the fits. 
The fitting procedure is quite stable: if an exception occurs the analysis has been automated to adjust the fitting parameters and ranges. In case of persistent anomalous behavior a flag is raised, the critical design point purged from the population and examined.  

%A two-level parallelization is implemented to speed-up the computation time as explained in Sec.~\ref{subsec:computing}.

%A comparison between the non-projective and projective designs of the inner tracker is also shown in Fig.~\ref{fig:tracking_projective}, where the projective design concentrates the material in a smaller dead area resulting in better resolution on a wider range of the pseudorapidity.
%

\begin{figure*}[!]
    \centering
    \includegraphics[width=0.245\textwidth]{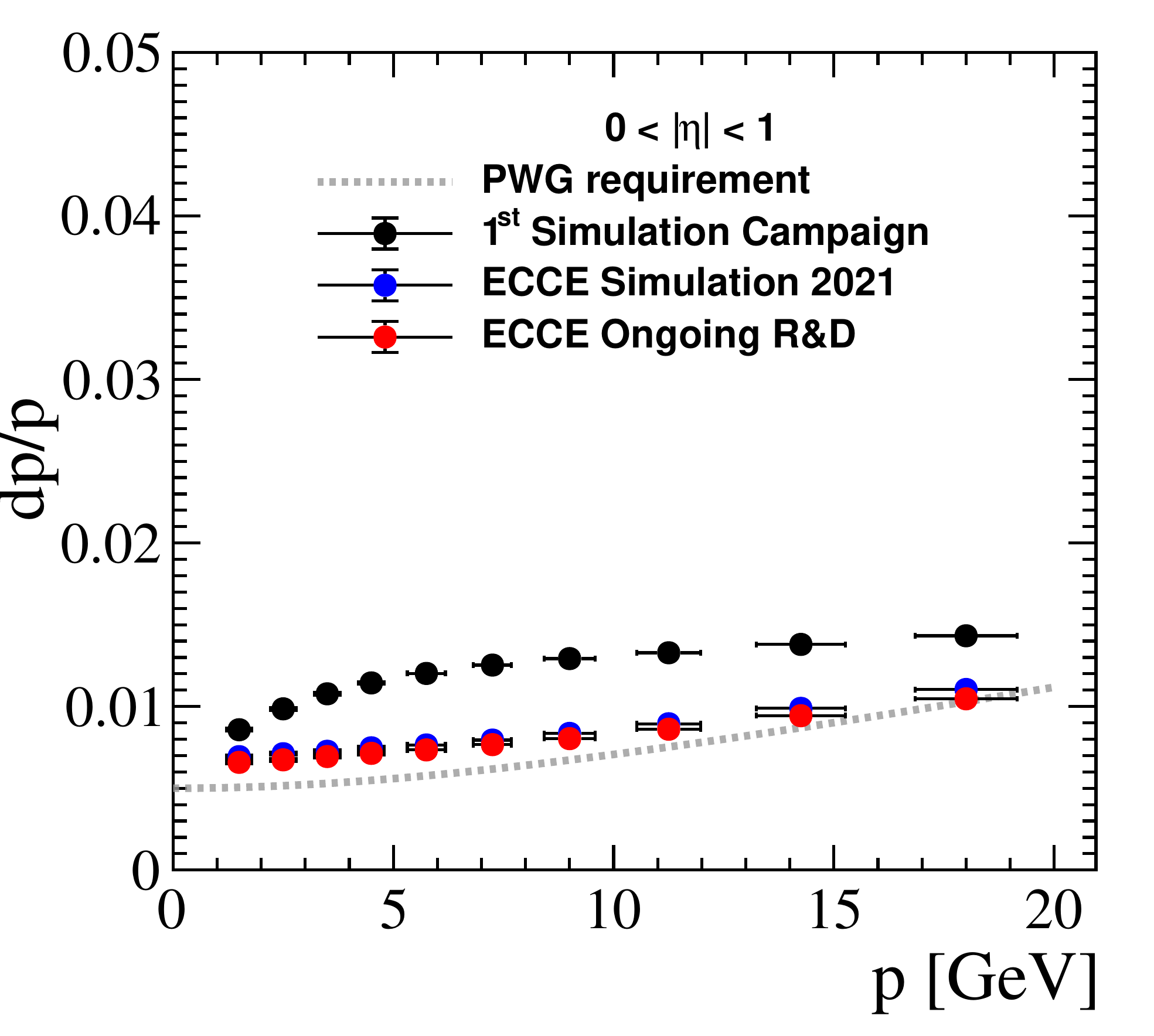}
    \includegraphics[width=0.245\textwidth]{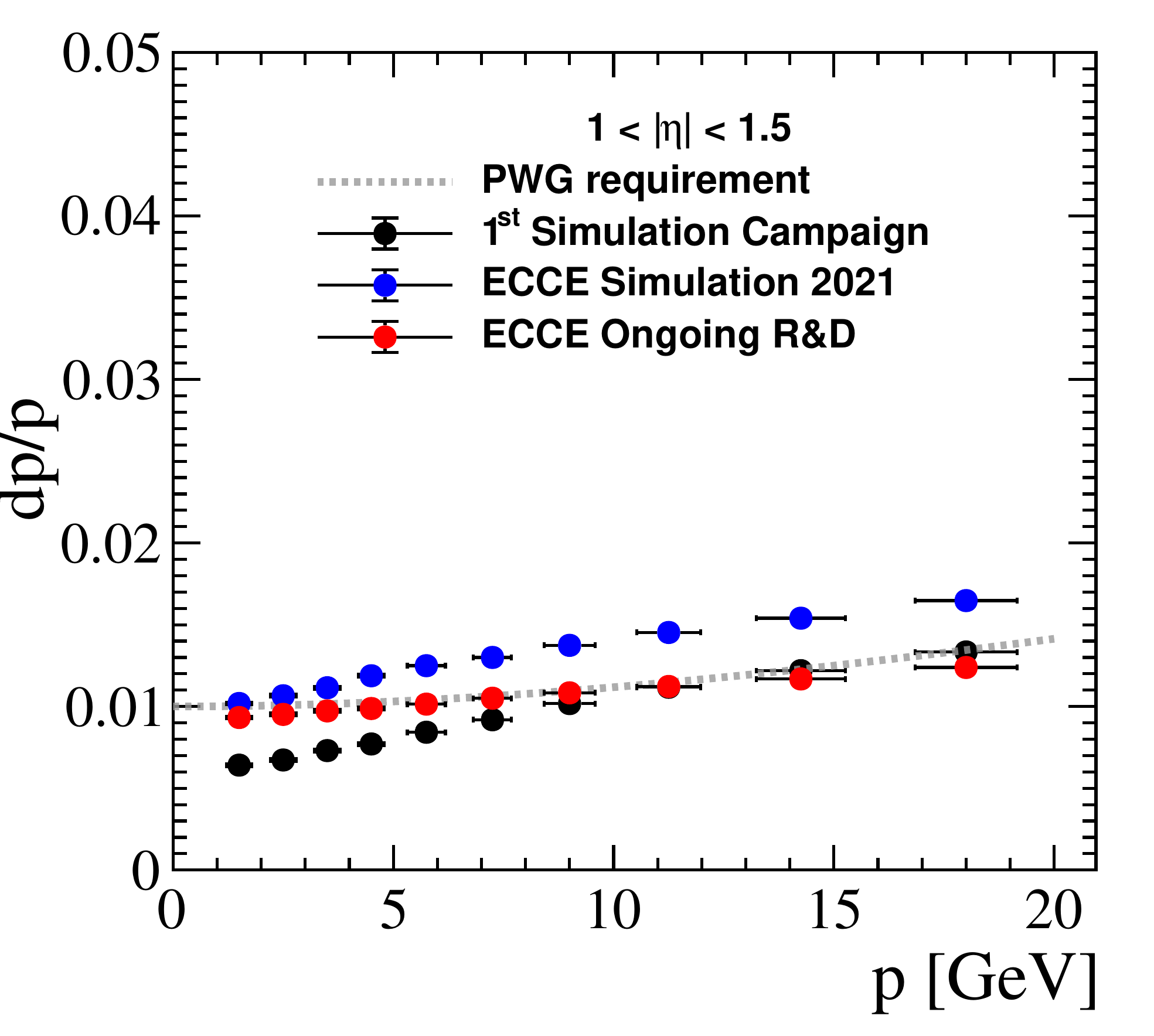}
    \includegraphics[width=0.245\textwidth]{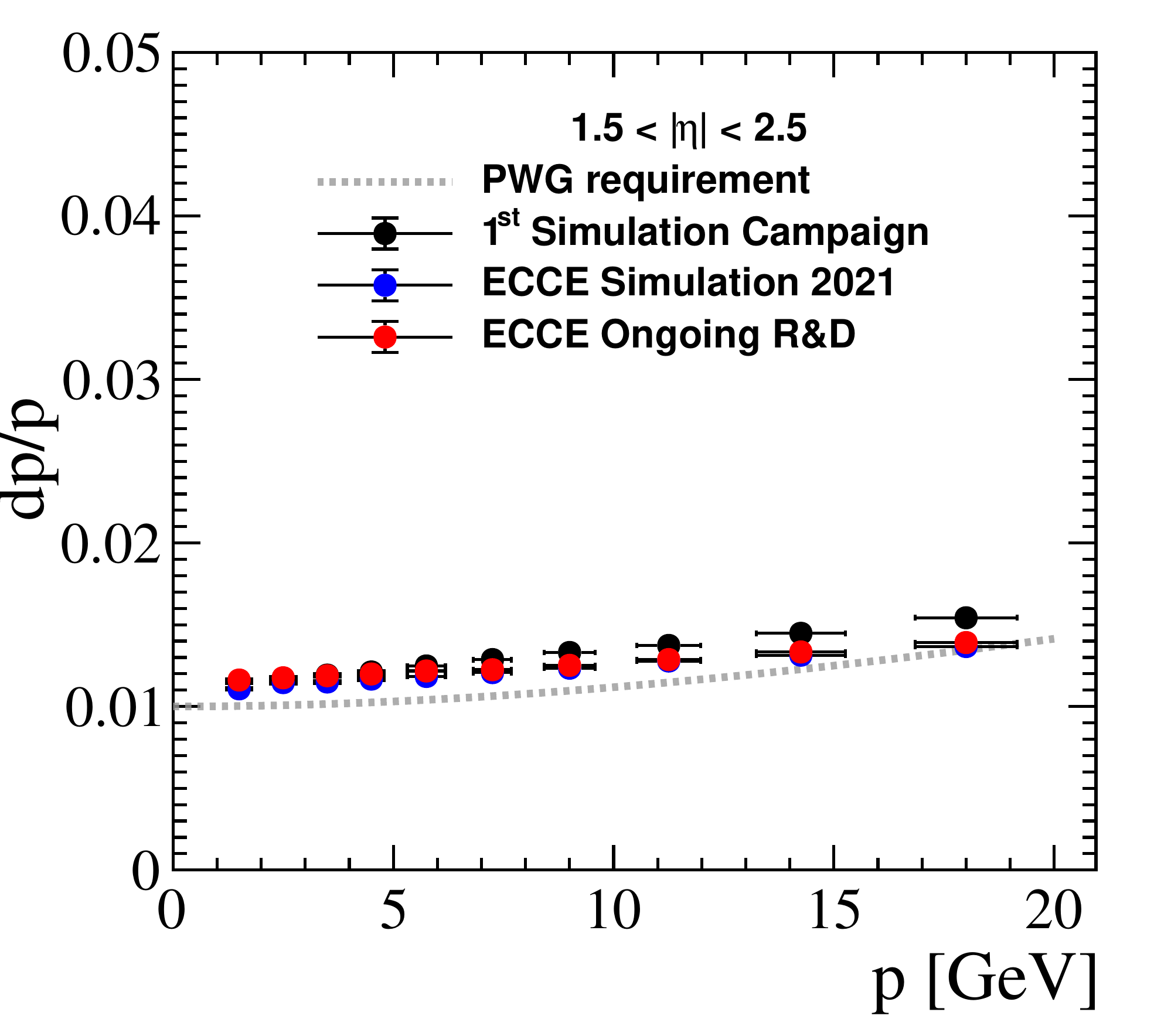}
    \includegraphics[width=0.245\textwidth]{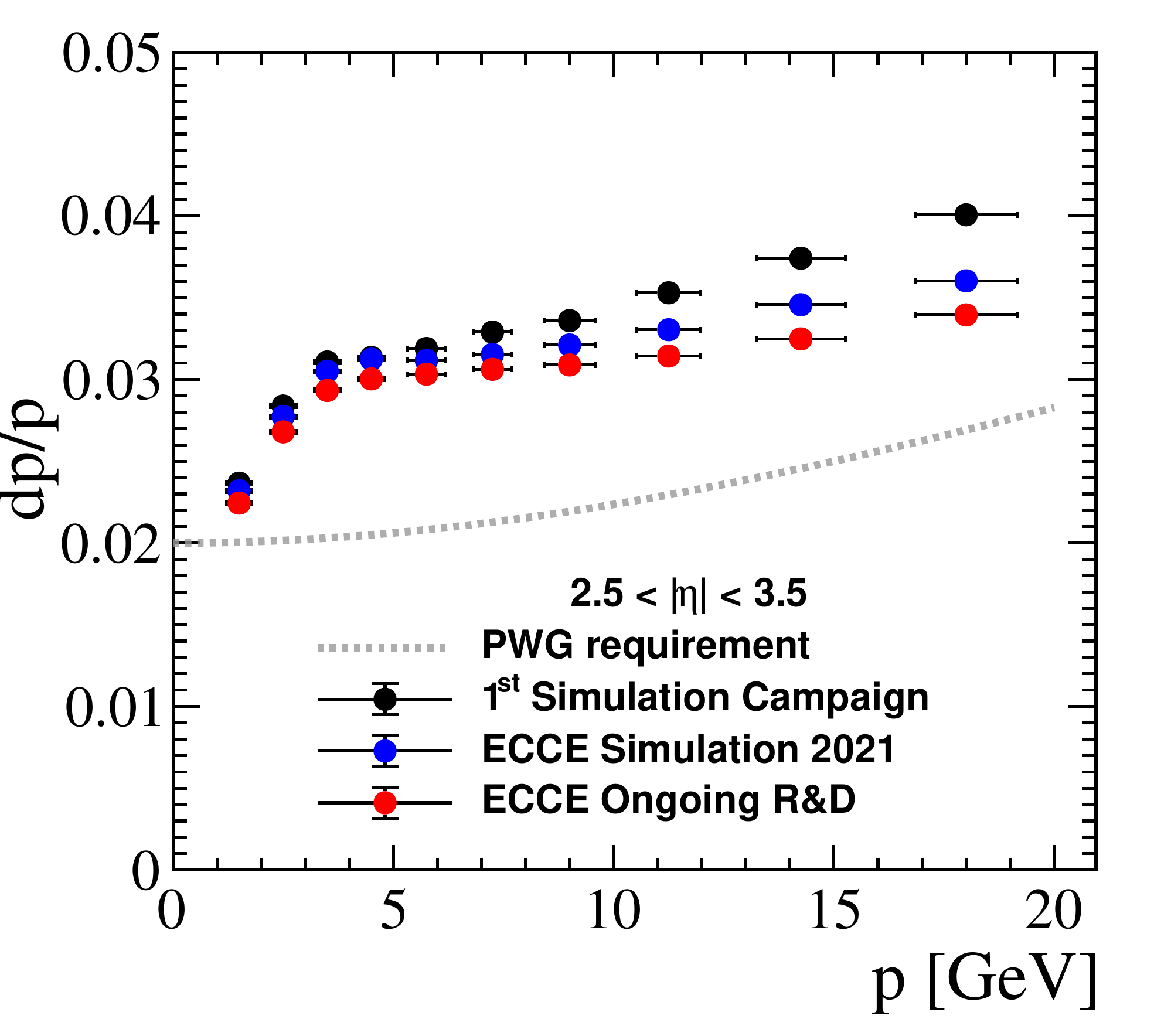}
    \caption{\textbf{Evolution of momentum resolution driven by the continued MOO:} Momentum resolutions in four bins of $\eta$. From left to right, respectively: $|\eta|<$ 1.0 corresponding to the barrel, 1.0 $|\leq \eta<|$ 1.5 corresponding to the transition region, and 1.5 $\leq |\eta| <$ 2.5 and 2.5 $\leq \eta <$ 3.5 corresponding to the two end-caps. Black points represent the first simulation campaign, and a preliminary detector concept in phase-I optimization which did not have a developed support structure; blue points represent the fully developed simulations for the final ECCE detector proposal concept; red points the ongoing R\&D for the optimization of the support structure. Compared to black, there is an improvement in performance in all $\eta$ bins with the exception of the transition region, an artifact that depends on the fact that black points do not include a realistic simulation of the material budget in the transition region. In the transition region, it can be also appreciate the improvement provided by the projective design. %\cris{polish text; reference in the document; move figures and reduce blank spaces}
    }
    \label{fig:continued_optimization}
\end{figure*}

The improvement obtained with the continual multi-objective optimization process is summarized in Fig. \ref{fig:continued_optimization}, where the momentum resolution obtained during phase-I optimization using a preliminary detector concept is compared to both the non-projective and the projective R\&D designs which are instead derived from fully developed simulations in phase-II optimization.

A detailed description of the optimized performance for all the objectives (momentum, angular resolutions and Kalman Filter efficiency) can be found in~\ref{app:comparison}.

%%%%%%%%%%%%%%%%%%%%%%%%%%%%%%%%%%%%%%%%%%%%%%%
\subsection{Physics Analysis}\label{subsec:physics_analysis}

To show a comparison in physics performance between the non-projective and projective designs, we analysed ${D}^{0}$ meson decay into $\pi^{+}K^{-}$. 
Data have been produced utilizing SIDIS events generated with Pythia6 \cite{Sjostrand:2006za},  corresponding to $ep$ events with 18 GeV $\times$ 275 GeV and high Q$^{2}$.\footnote{production: prop.5/prop.5.1;  generator: pythia6; kinematics: ep-18x275-q2-high. More info can be found in \cite{ecce_simulation_WG}.} 

\begin{figure}[!]
    \centering
    \includegraphics[width=0.48\textwidth]{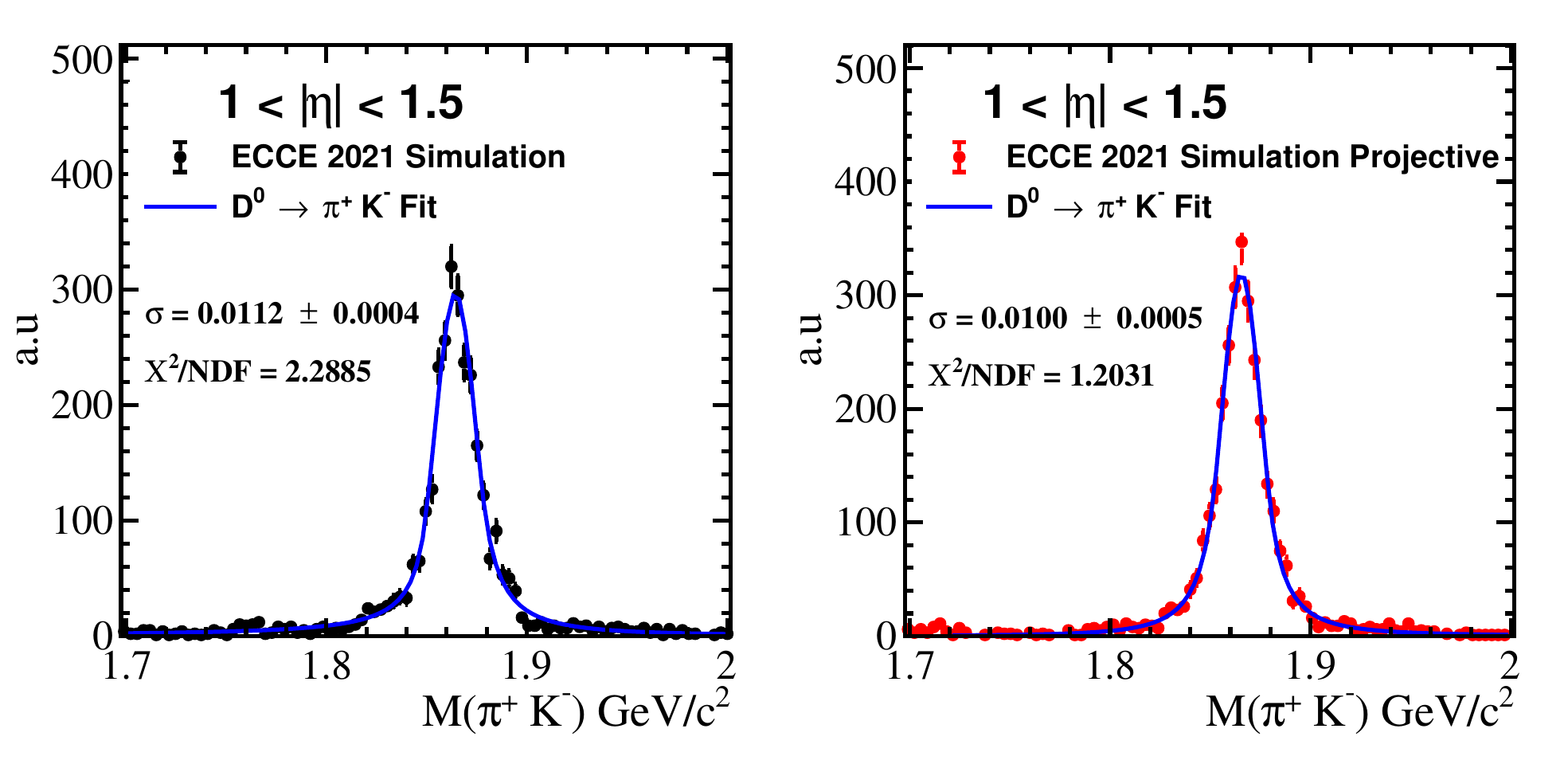} 
    \caption{\textbf{Analysis of SIDIS events with AI-assisted detector designs: } The $\pi^{+}K^{-}$ invariant mass obtained from the SIDIS events reconstructed in ECCE. A comparison of the resolution on the ${D}^{0}$ meson obtained with the non-projective and the projective design of the ECCE tracker is shown. 
    In order to emphasize the improvement in the region between $1.0 < |\eta| < 1.5$, a selection was made such that both the tracks $\pi^{+}$ or the $K^{-}$ are within the region of interest. The resolution with the projective design is improved by more than 10\% relative to the resolution obtained with the non-projective design. 
    \label{fig:physics_analysis_DSCB}
    }
\end{figure}

In Fig.~\ref{fig:physics_analysis_DSCB}, the $\pi^{+}K^{-}$ invariant mass is fitted with a double-sided Crystal Ball function \cite{aaboud2016search,lhcb2020first}. 

The decay events are selected in such a way to have at least one particle (either $\pi$ or $K$, or both) in the pseudorapidity bin 1.0$<$$|\eta|$$<$1.5, where the projective design is expected to improve the performance by concentrating all the material in a smaller dead area compared to the non-projective design. 

The analysis shows that the resolution obtained with the projective design is improved by more than 10\% relative to that obtained with the non-projective design. 
We also calculate the efficiency, defined as the number of reconstructed ${D}^{0}$ mesons divided by the number of true ${D}^{0}$ mesons. 
The efficiency obtained with the two designs  is consistent within the statistical uncertainties.

\section{Computing Resources}\label{sec:computing}

\begin{figure}[!]
    \centering
    \includegraphics[width=0.48\textwidth]{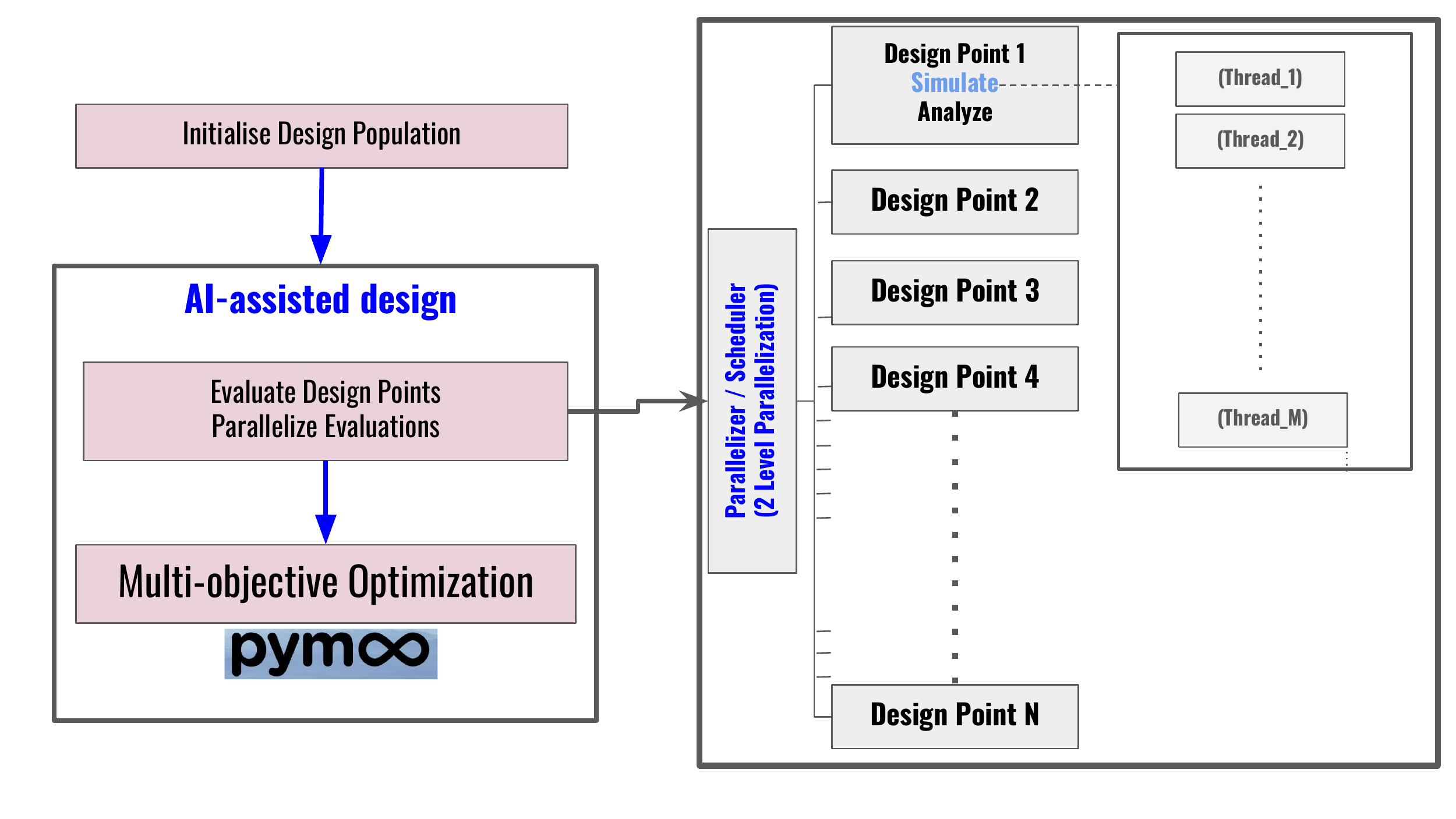}
    \caption{\textbf{Flowchart of the two-level parallelization implemented for detector design:}  Multiple design points are produced in parallel, and each design point is parallelized. The design is assisted by an AI-strategy, in this case MOO with evolutionary algorithms supported by pymoo \cite{blank2020pymoo}.
    \label{fig:parallelization}
    }
\end{figure}

%%%%%%%%%%%%%%%%%%%%%%%%%%%%%%%%%%%%%%%%
%\subsection{Computing Resources}\label{subsec:computing}
\paragraph{Parallelization}
A two-level parallelization has been implemented in the MOO framework: the first level creates the parallel simulations of design points, the second level parallelizes each design point (see Fig.~\ref{fig:parallelization}). 
The evaluation itself can be distributed to several workers or a whole cluster with libraries like Dask \cite{dask}.

\begin{figure}[h!]
    \centering
    \includegraphics[trim=7.5em 0em 7.8em 0em, clip, width=0.45\textwidth]{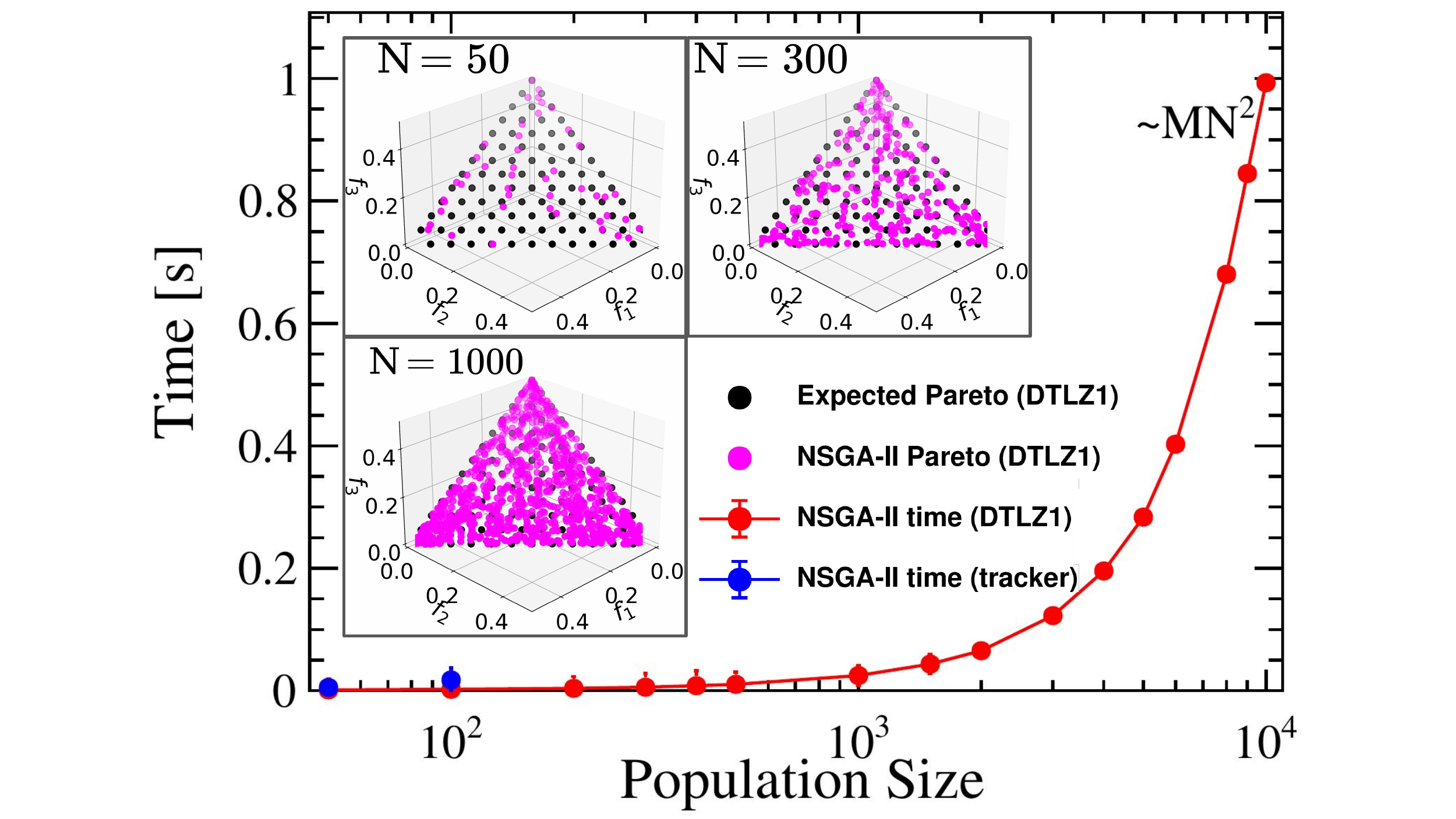}
    \caption{\textbf{The computing time taken by the genetic algorithm and the sorting in NSGA-II:}. Performance has been benchmarked with test problems like DTLZ1 \cite{ishibuchi2016performance} (red points) and the scaling $\sim MN^{2}$ (see Table~\ref{tab:hyperparameters}) has been verified with convergence to the Pareto front. The simulation time dominates the AI time during the optimization process (blue point). A two-level parallelization has been introduced in the framework to reduce this bottleneck.  
    The AI contribution typically becomes dominant when very large population size is needed to accurately approximate the Pareto front (cf.\cite{liu2020parallelization}). 
}
    \label{fig:computing-times}
\end{figure}

%%%%%%%%%%%%%%%%%%%%%%%%%%%%%%%%%%%%%%%%
\paragraph{Computing Budget}
Computing time studies have been carried out to evaluate the simulation time of each single design point as a function of the number of tracks generated. 
We made this study with simulations that included the tracking system and the PID system and estimated an effective simulation time of 0.2 s/track after removing an initial latency time. 
Similarly we made studies of the computing time taken by the AI-based algorithm in generating a new population of design points. 
Results of these studies are summarized in Fig.~\ref{fig:computing-times}.  

A larger population allows to approximate the Pareto front with larger accuracy. Extension of the design parameter space and the objective space to larger dimensionality implies a larger amount of CPU time which is mainly dominated by simulations if the population size remains smaller than 10$^{4}$--10$^{5}$, see Fig.~\ref{fig:computing-times}. 
%An interesting trend in recent years when the sorting time becomes dominant compared to simulations is that of implementing evolutionary MOO in supercomputers (see, \textit{e.g.}, \cite{liu2020parallelization}). 

For our goals the optimization pipelines of the ECCE tracking system were parametrized with 10--20 design parameters and 3--4 objectives; this allowed us to achieve good convergence with evolutionary MOO using a two-level parallelization strategy, and deployment on single nodes of 128 CPU cores available on the sci-comp farm at Jefferson Lab \cite{scicomp}.\footnote{Work is in progress to efficiently distribute the optimization pipeline to multiple nodes.}.

%\paragraph{Estimated Resources}

\paragraph{Planned Activities}
As described in this document, detector optimization with AI is an essential part of the R\&D and design process and it is anticipated to continue after the detector proposal.
%
%AI assisted in steering the design of the ECCE Tracker and can be utilized to fine-tune its design.
%Activities in this regard are anticipated to further extend the parametrization of the ECCE tracker. 
%
The AI-assisted design optimization of the ECCE inner tracker was based on evolutionary algorithms. During the detector proposal multiple optimization pipelines were run each with a population size of 100, representing different detector design configurations. At each iteration, AI updated the population. The total computing budget for an individual pipeline amounted to approximately 10k CPU-core hours. 
This number depends on the dimensionality of the problem.
Larger populations may need to be simulated to cope with the increased complexity in order to improve the accuracy of the approximated Pareto front. Different AI-based strategies will be compared. 

Activities are planned to continue the detector optimization: new optimization pipelines can deal with a larger parameter space to include a system of sub-detectors such %as the whole ECCE tracker 
 or to combine tracking and PID in the optimization process. We also plan to optimize other sub-detectors like, \textit{e.g.}, the dRICH, leveraging on the expertise internal to the ECCE collaboration regarding specifically the design of the dRICH with AI-based techniques \cite{cisbani2020ai}. 
%
%For the tracker we used as driving criteria figures of merit that embody the performance of the tracking detector: momentum and angular resolution, Kalman filter efficiency. 
%After the detector optimization, we validate the performance on other figures of merit not utilized during the optimization process.
%We also completed these studies by analysing the performance of the optimized detector on physics channels, particularly we made an analysis of the $D^{0}\rightarrow \pi^{+}K^{-}$ that showed an improved resolution by at least 10\% with the projective tracker design as compared to the non-projective geometry. 
As a future activity we aim to encode physics-driven objectives in the MOO problem. 
A thorough comparison of results obtained with different AI-based strategies (\textit{e.g.}, MOO based on genetic algorithms or bayesian approaches) can be also studied. 

We anticipate for 2022 roughly 1M CPU-core hours for these activities.

\section{Summary}\label{sec:summary}

%EIC has been successfully using AI to assist the design optimization of the detector \cite{cisbani2020ai}. 

Large scale experiments in high energy nuclear physics entail unprecedented computational challenges and the optimization of their complex detector systems can benefit from AI-based strategies \cite{fanelli2022design}.

In this paper we described the successful implementation of a multi-objective optimization approach to steer the multidimensional design of the ECCE tracking system, taking into account the constraints from the global detector design. 
This work was accomplished during the EIC detector proposal, and was characterized by a continued optimization process where multiple optimization pipelines integrating different configurations of sub-detectors were compared using full \geant simulations. 
The insights provided by AI in such a multi-dimensional objective space characterizing the detector performance (\textit{e.g.}, tracking efficiency, momentum and angular resolutions), combined to other aspects like risk mitigation and costs reduction, helped selecting the candidate technology of the ECCE tracker. 
This approach is being used in an ongoing R\&D project where the design parametrization has been extended to include the support structure of the tracking system. 

The design optimization can be also extended to tune the parameters of a larger system of sub-detectors. Physics analyses are at the moment done after the optimization for a given detector design solution candidate, but they can be encoded during the optimization process as physics-driven objectives in addition to objectives representing the detector performance. %A thorough comparison of results obtained with different AI-based strategies (e.g., multi-objective optimization based on genetic algorithms and bayesian multi-objective optimization) can be also studied. 

Detector optimization with AI is anticipated to continue after the detector proposal, and activities are planned to further optimize the tracking system, including PID sub-detectors, particularly the dual-RICH \cite{cisbani2020ai}.

%MOO has been used to ste

%These AI-based strategies can help to further optimize the proposed detector concept based on the experience of ECCE during the detector proposal.
\section*{Acknowledgements}
\label{acknowledgements}

We thank the EIC Silicon Consortium for cost estimate methodologies concerning silicon tracking systems, technical discussions, and comments.  We acknowledge the important prior work of projects eRD16, eRD18, and eRD25 concerning research and development of MAPS silicon tracking technologies.

We thank the EIC LGAD Consortium for technical discussions and acknowledge the prior work of project eRD112.

We thank (list of individuals who are not coauthors) for their useful discussions, advice, and comments.

We acknowledge support from the Office of Nuclear Physics in the Office of Science in the Department of Energy, the National Science Foundation, and the Los Alamos National Laboratory Laboratory Directed Research and Development (LDRD) 20200022DR.
%

%\input{sections/acknowledgements}
%% \appendix

%\clearpage
\bibliographystyle{elsarticle-num} 
%\bibliography{refs.bib,refs-notes.bib}
\bibliography{refs.bib,refs-ecce.bib}

%\clearpage
\appendix
\section{Details on Parametrization}\label{app:par_tracking}

%%%%%%%%%%%%%%%%%%%%%%%%%%%%%%%%%%%%%%%
\subsection*{Tracking System Parametrization} 
%%%%%%%%%%%%%%%%%%%%%%%%%%%%%%%%%%%%%%%

%\cris{Add references for each detector? Reference ECCE detector paper.}

\paragraph{Vertex layers} 
There are three vertex barrel layers in the ECCE tracking system made of MAPS technology. % (ITS3).  already defined
The vertex cylinder consists of strips which are made of pixels, where the individual sensor unit cell size is 17.8 mm $\times$ 30.0 mm.
The length of the vertex layers is fixed at 27 cm; %\cris{(30) drop?}
the radii of the three vertex layers are fixed to 3.4, 5.67, 7.93 cm, respectively.
For the non-projective design, the angle of the support structure with respect to the interaction point is fixed ($\theta = 36.5\degree$) and the radius of the support is at 6.3 cm, while the length of it is 17 cm.
For the projective design, the radius of the support structure is the same, while the length is calculated based on the angle of projection and the radius as shown in Fig.~\ref{fig:SupportStructureparametrization}.

\paragraph{Sagitta layers} 
There are two sagitta barrel layers in the ECCE tracking system.
The sagitta barrel layers are made of MAPS technology and have 
%
%The sagitta barrel layer has a 
fixed length of 54 cm.
For the non-projective design the radii of the sagitta layers are 21.0, 22.68 cm, respectively. For the projective parameterization,the radius of the sagitta barrel is calculated such that there are no gaps in the acceptance of the region enclosed by the barrels, according to the following equation:
%. The radius of the barrel layer is calculated as: 
$$r_{sagitta} = \frac{l_{sagitta}}{2}\tan{\theta}.$$

The radius of the sagitta layers is also constrained since the strips have fixed width  $w$ = 17.8 mm; therefore we want to minimize the quantity: 
%
%$$\frac{2 \pi r_{sagitta}}{1.78} = n; n \in \mathbb{Z}^{+}.$$
%
$$ min \Bigg\{ \Bigg| \frac{2 \pi r_{sagitta}}{w} - \floor*{\frac{2 \pi r_{sagitta}}{w}} \Bigg| \Bigg\},$$
 where $\floor*{x}$ represents the ceiling of x.

\paragraph{$\mu$Rwell layers}
In the ECCE tracking system there are three cylindrical $\mu$Rwell layers, each %end of each %$\mu$Rwell barrel layers.
 endowed with a support ring.  
An extended supporting plateau is included at either ends of the $\mu$Rwell to rest the entire cylindrical detector on this platform. 
This results in a constant shift of the support cone by the plateau length (5 cm) as shown in Fig.~ \ref{fig:SupportStructureparametrization}.
%
%Therefore, for each $\mu$Rwell layer the support cone is shifted constantly by the length of the plateau.
%
For both the non-projective design and the projective design the $\mu$Rwell-1 radius is a free parameter. The length of the $\mu$Rwell-1 is calculated based on the angle of the conical support structure. 
In the non-projective design we have the conical support structure angle fixed ($\theta = 36.5\degree$), therefore the length of $\mu$Rwell-1 depends only on its radius; $\mu$Rwell-2 has its radius as a free parameter;
since the angle of the conical support structure is fixed the length of $\mu$Rwell-2 depends on its radius. 
%The optimization always preferred to maximize the radius and therefore was fixed in the projective design.  
%
 In the projective design instead the $\mu$Rwell-2 has a fixed radius of 51 cm ($i.e.$, $r_{max} +$ 1 cm). The length of the $\mu$Rwell-2 is calculated based on the angle of the conical support structure. The length of the $\mu$Rwell takes into account the constant shift due to the plateau.  
The dimensions of $\mu$Rwell-3 are fixed in both non-projective and projective designs; the $\mu$Rwell-3 is outside of the inner tracking system and it has radius of 77 cm and a total length is 290 cm.

\paragraph{EST/FST disks}
%

%The FST/EST disks are made of MAPS technology (ITS3). 
%The disks are tiled up using MAPS pixel which comes in fixed dimension of 17.8 mm $\times $ 30.0 mm; we already said this
%
For both the non-projective and projective designs, 
$R_{min}$ of the disks must be compatible with the beam pipe envelope which increases in radius as a function of $z$; 
$R_{out}$ of the disks is parametrized to be compatible with the support cone structure shown in Fig. \ref{fig:SupportStructureparametrization} which has an angle $\theta$ that is variable in the projective design and fixed in the non-projective case. 
For the non-projective design, the $z$ positions of the disks were all free parameters in the first optimization pipelines. However, to maximize the hit efficiency, some disks have been eventually placed within the support cone at the beginning of every plateau (Fig. \ref{fig:SupportStructureparametrization} with fixed angle $\theta = 36.5\degree$). Therefore, two disks in the electron-going direction and two disks in the hadron-going direction are not free to vary in $z$. 
For instance, consider Fig.~\ref{fig:ECCE_Tracking_non_proj_and_proj} (right), where EST3, EST4, FST3, FST4 are placed at the begin of the pleateau, whereas the disks EST1, EST2, FST1, FST2, FST5 are free to vary in $z$ position.
The same parameterization is extended to the projective design and made compatible with a varying conical support structure.

As the disks are tiled up using MAPS pixels, the difference between $R_{min}$ and $R_{out}$ is constrained to optimize the sensor coverage for all disks; this is implemented by means of two functions, namely:
%
%$$(R_{max} - R_{min})/d = n; n \in \mathbb{Z}^{+},$$
%%$$(R_{max} - R_{min})/30.0 = n; n \in \mathbb{Z}^{+}$$
%
% where $d=$ 17.8 or 30.0 mm. 
%
% to account for the constraints in all the FST and EST disks in the end caps region, 
% this is done with two functions (one for $d=$ 17.8,  30.0 mm) are considered, namely:
 $$min \Bigg\{ \sum_{i}^{all \ disks} \Bigg| \frac{R^{i}_{out} - R^{i}_{in}}{d} - \floor*{\frac{R^{i}_{out} - R^{i}_{in}}{d}} \Bigg| \Bigg\},$$
%$$\sum_{i} \Bigg| \frac{R^{i}_{max} - R^{i}_{min}}{30.0} - \text{ int}\Bigg(\frac{R^{i}_{max} - R^{i}_{min}}{30.0}\Bigg) \Bigg| == 0$$
%
%where $\floor*{x}$ represents the ceiling of x. 
 where $d=$ 17.8, and 30.0 mm. 
 This limits the amount of violation made by a design solution. 
%\cris{We should add this is done by adding a constraint function. @karthik: can you add?}

\paragraph{TOF system} 
The central barrel TOF (CTTL) is an AC-LGAD based TOF detector with a fixed radius of 64 cm and a fixed length of 280 cm.
The TOFs at the electron-going endcap (ETTL) and the hadron-going endcap (FTLL) are AC-LGAD-based TOF disks.  
For the non-projective design the TOF detectors have fixed dimensions. 
For the projective design the TOF detectors in the end cap regions have their $z$ positions as free parameters. 
$R_{in}$ and $R_{out}$ of the ETTL/FTTL disks depends on the position of the disk $z$. The $R_{in}$ of the disk should be compatible with the radius of the beam envelope which increases linearly as a function of $z$; $R_{out}$ of the disks varies as a function of $z$ such that the acceptance coverage by the ETTL/FTTL is roughly unaltered.

\begin{comment}
\paragraph{MPGD System \cris{keep here?}}
%
EMPGD and FMPGD are disk-like gaseous detectors placed in the electron-going and in the hadron-going directions, respectively;
%
$R_{min}$ and $R_{max}$ of the EMPGD/FMPGD disk depends on the position of the disk $z$. The $R_{min}$ of the disk must be compatible with the radius of the beam envelope which increases linearly as a function of $z$.
%
$R_{max}$ of the disk varies as a function of $z$ such that the acceptance coverage provided by the EMPGD/FMPGD is roughly unaltered.
\end{comment}

\paragraph{PID Detectors}
The Detection for Internally Reflected Cherenkov light (DIRC) is a detector for PID in the barrel region. DIRC system has fixed dimensions and occupies a radial space from 71.5 cm to 76.6 cm. 
%, see Fig.~\ref{fig:ECCE_TrackingPID_Detailed}.
%
The modular RICH (mRICH) is a Ring Imaging Cherenkov detector system in the e-going direction with fixed dimensions. 
mRICH has a $z$ position starting at -135 cm extending in $z$ to -161 cm.
The dual Radiator Imaging CHerenkov (dRICH) detector is a detector system in the forward direction with fixed dimensions. dRICH has a $z$ position starting at 180 cm and extends up to 280 cm.
\newline
\newline
The thickness of the detectors and support structures are also taken into account to avoid overlaps between the detectors. 
The most recent optimization pipelines were extended to also include in the parametrization the outer tracking layers in the two endcaps, as explained in Sec. \ref{subsec:encoding}.    
An overlap check is performed each time a new design point is evaluated during the optimizaton process.

%%%%%%%%%%%%%%%%%%%%%%%%%%%%%%%%%%%%%%%
\subsection*{Support Structure Parametrization}\label{subsec:par_tracking}
%%%%%%%%%%%%%%%%%%%%%%%%%%%%%%%%%%%%%%%

\begin{figure}[h!]
    \centering
    \includegraphics[width=0.45\textwidth]{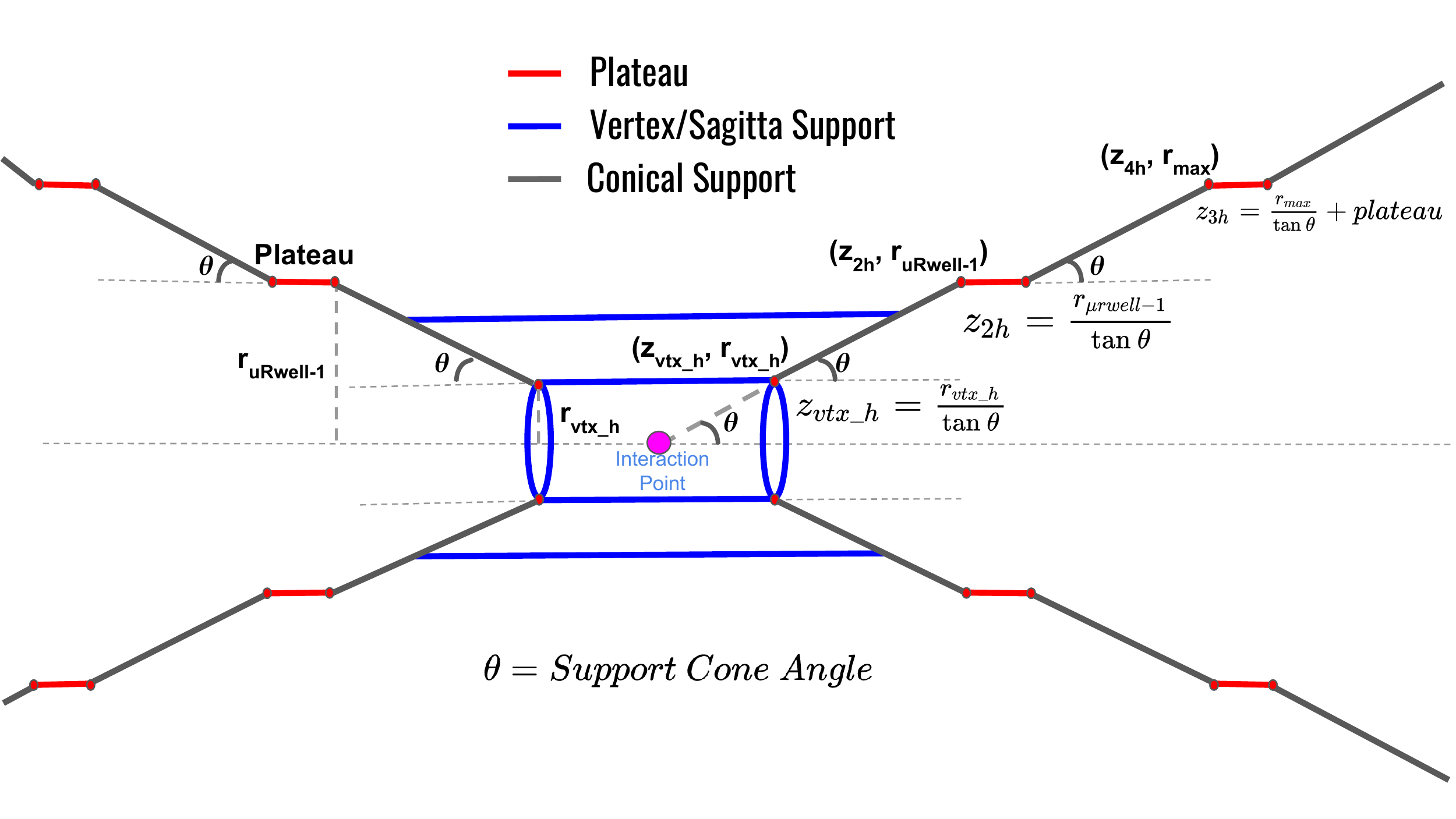}
    \caption{\textbf{Parametrization of the support cone for the inner tracker:} The inner tracker support is characterised by five variables: $\theta$ (the angle of projection of the support cone structure), $r_{vtx}$ (radius of vertex support structure), $r_{\mu rwell-1}$ $\mu$Rwell-1 radius, plateau length, $r_{max}$ maximum allowed radius of inner tracker).
    %\cris{$\mu$RWELL -1: penultimate? Does not appear in figure dubbed this way.}
    %\textbf{It is uRwell-1 is penultimate. }
    }
    \label{fig:SupportStructureparametrization}
\end{figure}

The implementation of the projective geometry of the inner tracker is described in Fig.~\ref{fig:SupportStructureparametrization}, which shows the parametrization used for the support cone structure of the inner tracker. 
Some parameters have been considered fixed and other free to vary within their ranges.
Parameters that are fixed typically do not have much room for optimization considering the constraints of the design and potential overlaps. 
The non-projective design can be realised by fixing the support structure angle to ($\theta = 36.5\degree$) shown in \ref{fig:SupportStructureparametrization}. Therefore, the non-projective design solutions are a subset of solutions that can be achieved by this parameterization.  

\section{Baseline and R\&D designs}\label{app:comparison}

%%%%%%%%%%%%%%%%%%%%%%%%%%%%%%%%%%%%%%%
\subsection*{Resolutions and Efficiency}
%%%%%%%%%%%%%%%%%%%%%%%%%%%%%%%%%%%%%%%

\begin{figure*}[h!]
    \centering
    \includegraphics[width=0.98\textwidth]{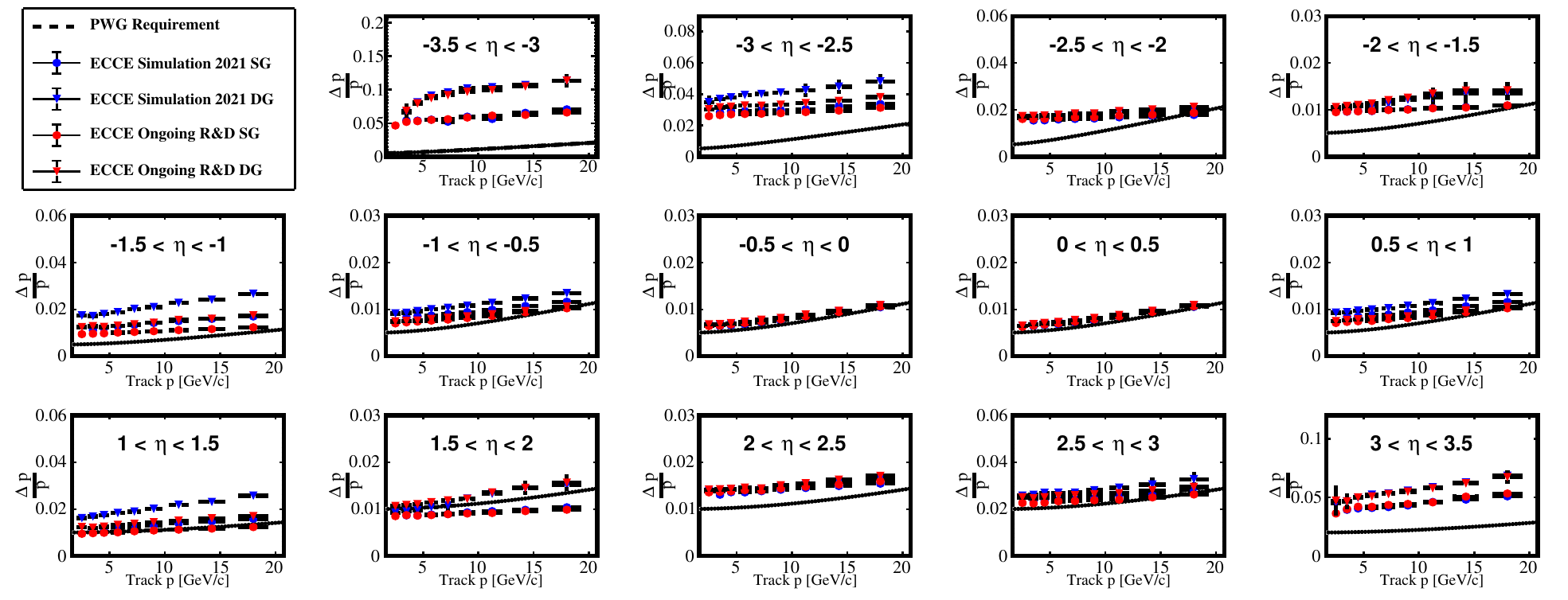}
    \caption{
    \textbf{Momentum resolution for the ECCE tracker reference design and the projective design (ongoing R\&D):} The $\Delta p /p$ are extracted with single-gaussian (SG) fits (circles) to be consistent with the YR PWG and compared to double-gaussian (DG) fits (triangles). DG (see Eq.~\eqref{eq:DG}) is used in the AI-assisted pipelines as it provides more stable fits. DG results are systematically larger than the SG ones as they capture more realistic distributions (see Fig.~\ref{fig:double_gauss}). 
    This effect is larger in $1<|\eta|<1.5$ for the non-projective design (blue points) since more material is traversed by the tracks; the same effect is moderately present also in $0.5<|\eta|<1$ for the same reason. 
    In the same bins, the ongoing R\&D design shows a significant reduction of the impact of the readout and services and SG (red circle) and DG (red triangles) are close to each other.  
    DG points are systematically large in $3<\eta<3.5$ and in $-3.5<\eta<3.0$, and this corresponds to edge effects at large $|\eta|$ values close to the inner radii of the disks in the endcaps.  
    In the region $-1<\eta<3.0$ $\Delta p/p$ is consistent with the YR physics requirements with SG. Note that physics in the backward region relies on the EM calorimeter also, and in the proposal it has been demonstrated that larger resolutions are acceptable.  
    \label{fig:Phase-II_Momentum_res}
    }
\end{figure*}

\begin{figure*}[!]
    \centering
    \includegraphics[width=0.98\textwidth]{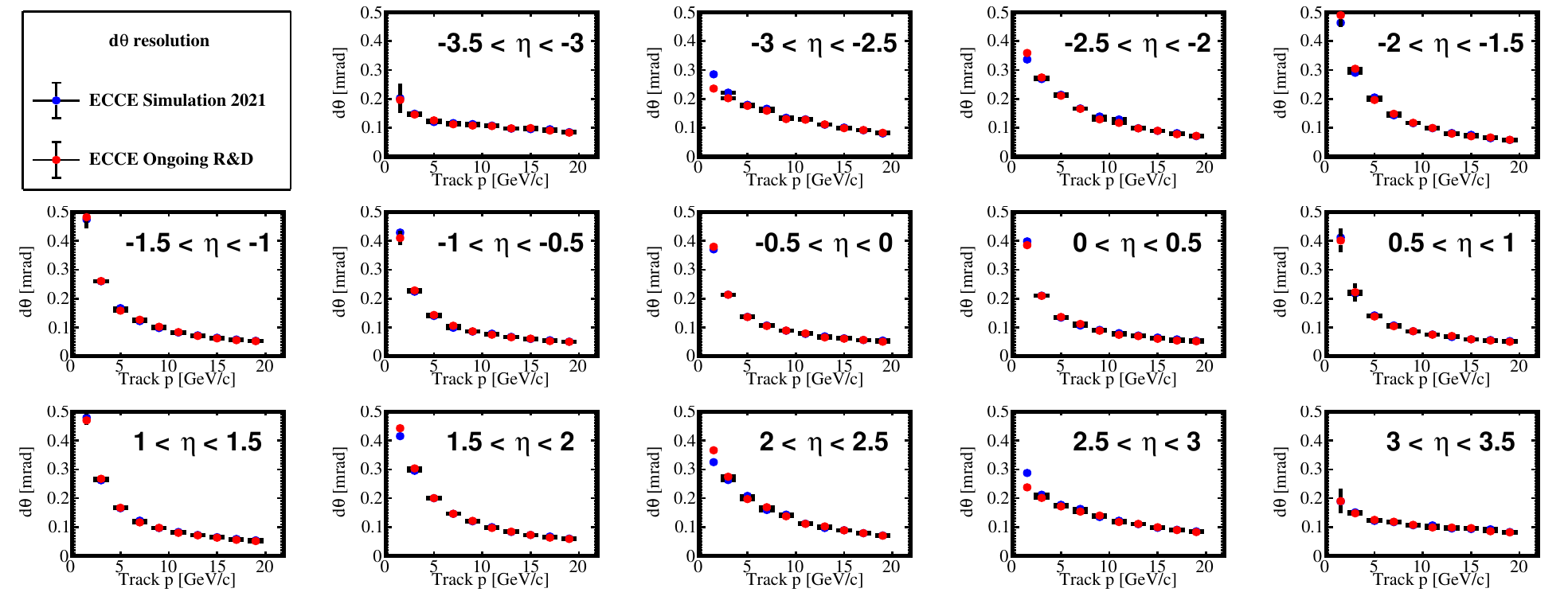}
    \caption{\textbf{Angular resolution ($d\theta$) for ECCE projective geometry and ECCE simulation:} The resolution shows an improvement in the transition region between $1.0 < |\eta| < 1.5$. These plots have been produced using double-Gaussian fits.
    \label{fig:Phase-II_dth_res}
    }
\end{figure*}

\begin{figure*}[!]
    \centering
    \includegraphics[width=0.98\textwidth]{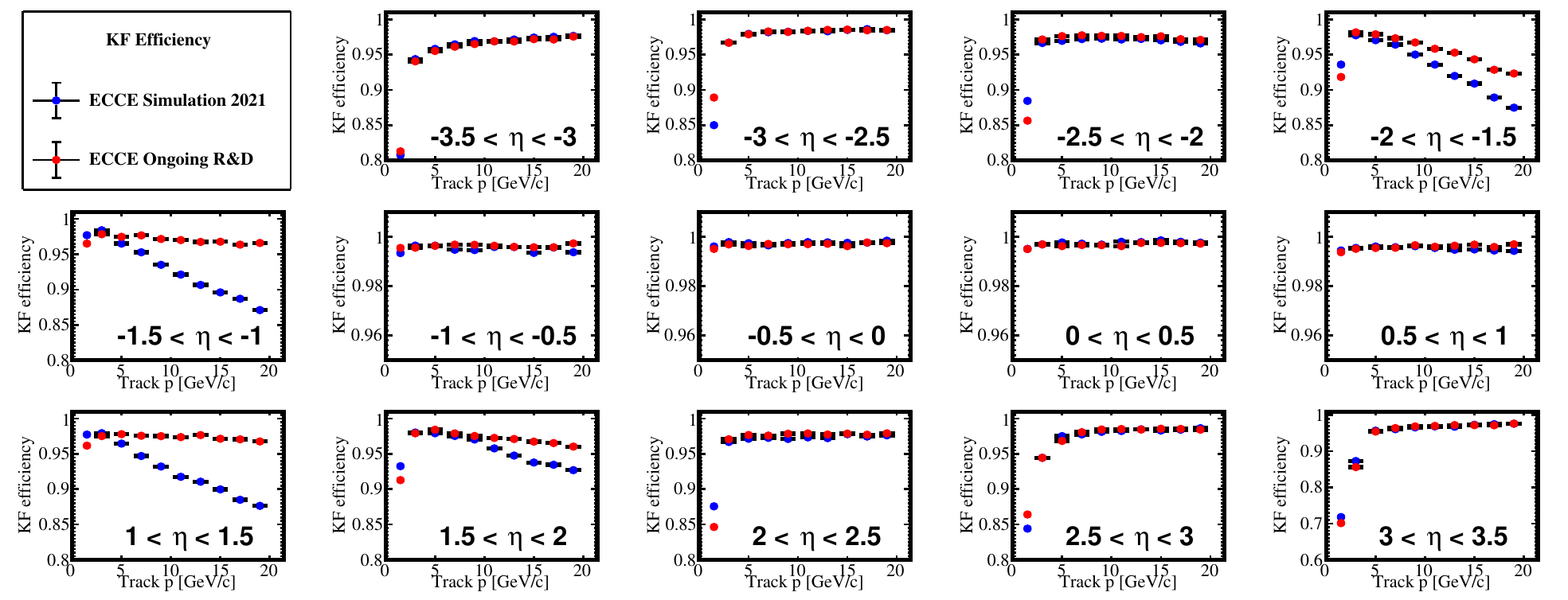}
    \caption{\textbf{KF Efficiency for ECCE projective geometry and ECCE simulation:} Reconstruction efficiency shows an improvement in the transition region between $1.0 < |\eta| < 1.5$. There is a significant drop in the transition region caused due to multiple scattering of the support structure. }
    \label{fig:Phase-II_KF_eff}
\end{figure*}

A thorough comparison between the non-projective ECCE simulation and the ongoing R\&D was carried out to optimize the support structure through a projective design.  
Fig.~\ref{fig:Phase-II_Momentum_res} shows a study of the obtained  momentum resolution 
%for the non-projective ECCE tracker reference design and the ongoing project R\&D on the projective mechanical structure for the ECCE tracker design 
in bins of pseudorapidity.
Similarly Fig.~\ref{fig:Phase-II_dth_res} shows the angular resolution while Fig.~\ref{fig:Phase-II_KF_eff} shows the Kalman filter efficiency as defined in Eq.~\eqref{eq:objective_KF}.

%\begin{comment}

%\end{comment}

%%%%%%%%%%%%%%%%%%%%%%%%%%%%%%%%%%%%%%%
\subsection*{Validation}\label{subsec:validation}
%%%%%%%%%%%%%%%%%%%%%%%%%%%%%%%%%%%%%%%

\begin{figure*}[!]
    \centering
    \includegraphics[width=0.98\textwidth]{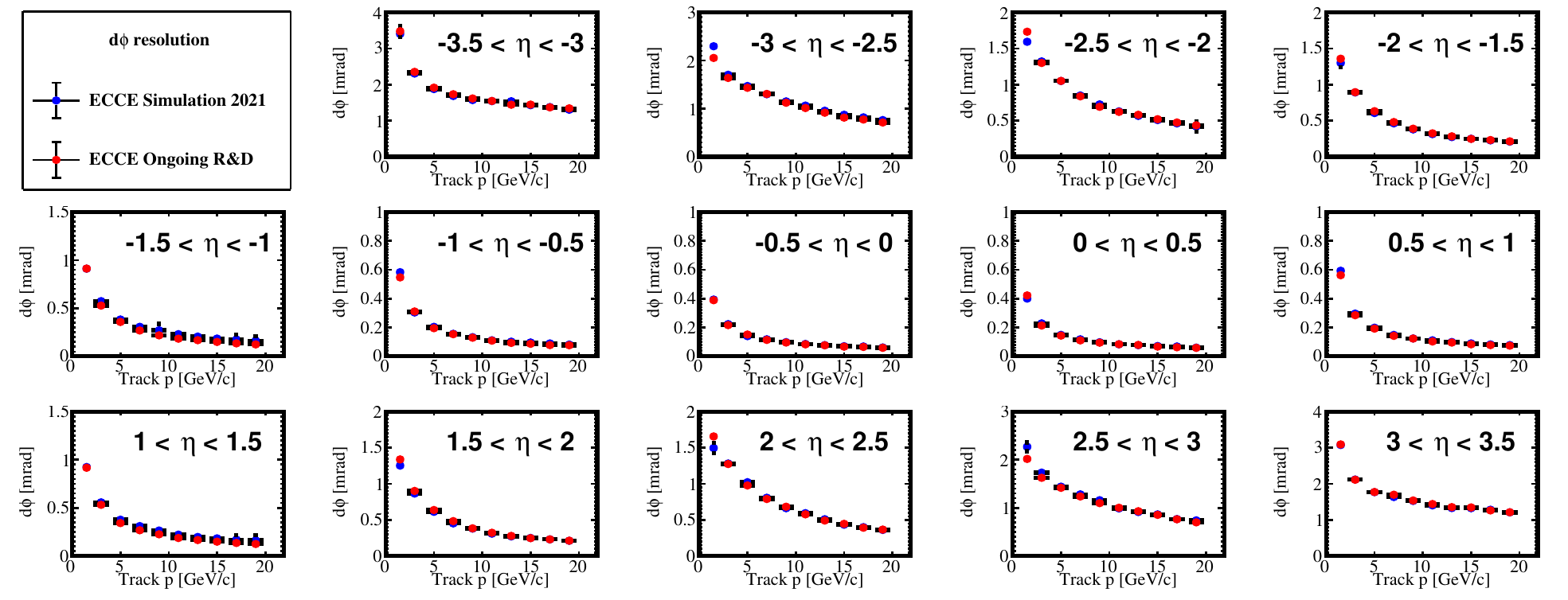}
    \caption{\textbf{Angular resolution ($d\phi$) for ECCE projective geometry and ECCE simulation:} The resolution shows an improvement in the transition region between $1.0 < |\eta| < 1.5$. These plots have been produced using double-Gaussian fits.}
    \label{fig:Phase-II_dph_res}
\end{figure*}

\begin{figure*}[!]
    \centering
    \includegraphics[width=0.98\textwidth]{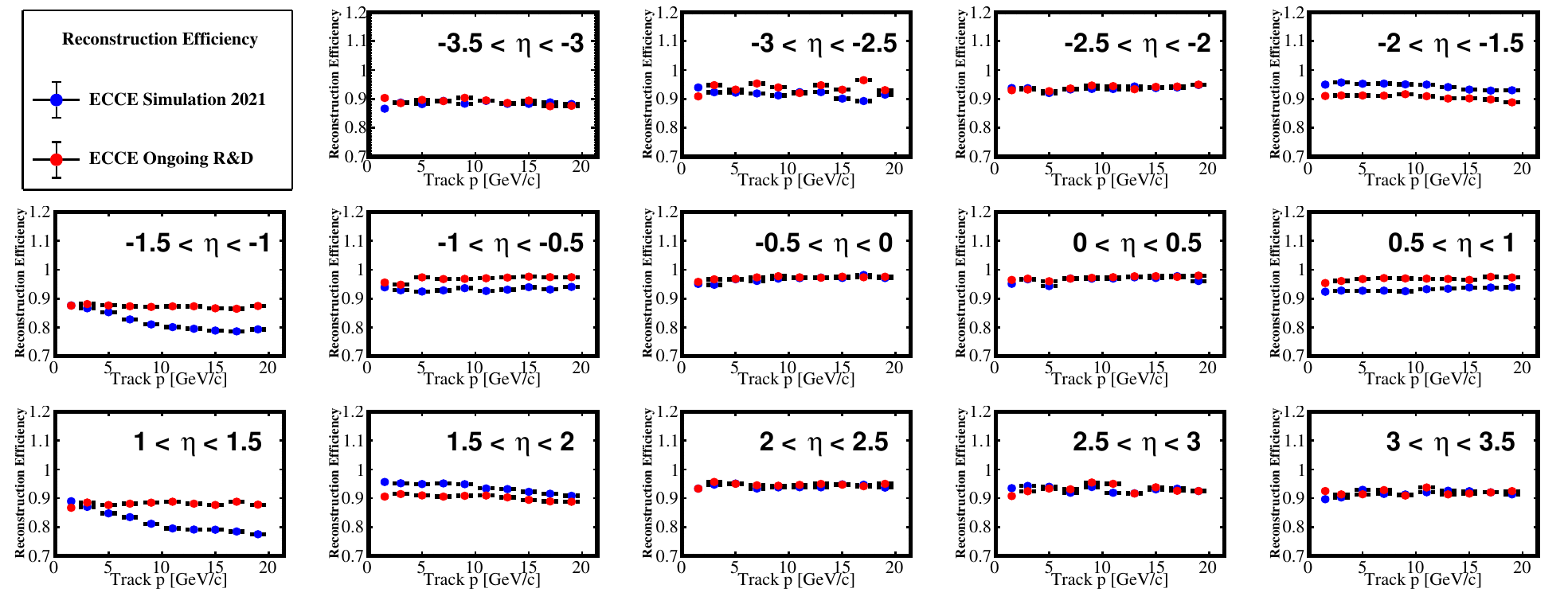}
    \caption{\textbf{Reconstruction efficiency for ECCE projective geometry and ECCE simulation:} Reconstruction efficiency shows an improvement in the transition region between $1.0 < |\eta| < 1.5$}
    \label{fig:Phase-II_recon_eff}
\end{figure*}

Validation is performed by looking at figures of merit that are not used during the optimization process. 
In Sec.~\ref{subsec:physics_analysis} we already described a physics analysis with SIDIS events that further consolidates our conclusions.
We show here additional examples of validation: Fig.~\ref{fig:Phase-II_dph_res} and Fig.~\ref{fig:Phase-II_recon_eff} display the azimuthal angular resolution and the reconstruction efficiency obtained for both the non-projective and the projective tracker designs.
The azimuthal resolution looks consistent within the uncertainty while the reconstruction efficiency looks in general better for the projective design, particularly in the 1 $<$ $|\eta|$ $<$ 1.5 region where the non-projective design has a larger dead area that affects the reconstruction of tracks.

%--------------
%\begin{comment}

%\end{comment}

\begin{figure*}[!]
\centering
    \centering
    \includegraphics[width=0.3\textwidth]{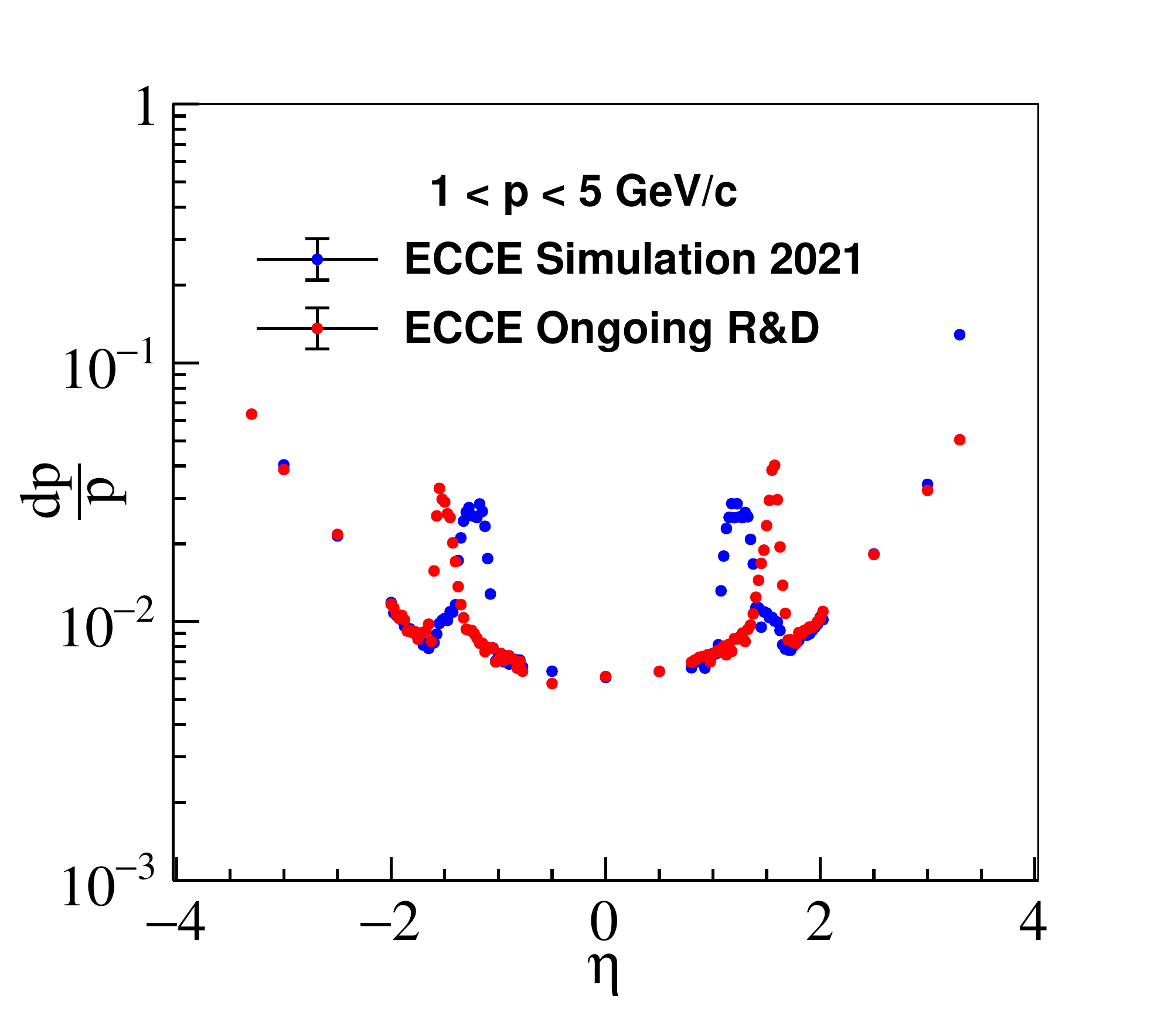}
    \includegraphics[width=0.3\textwidth]{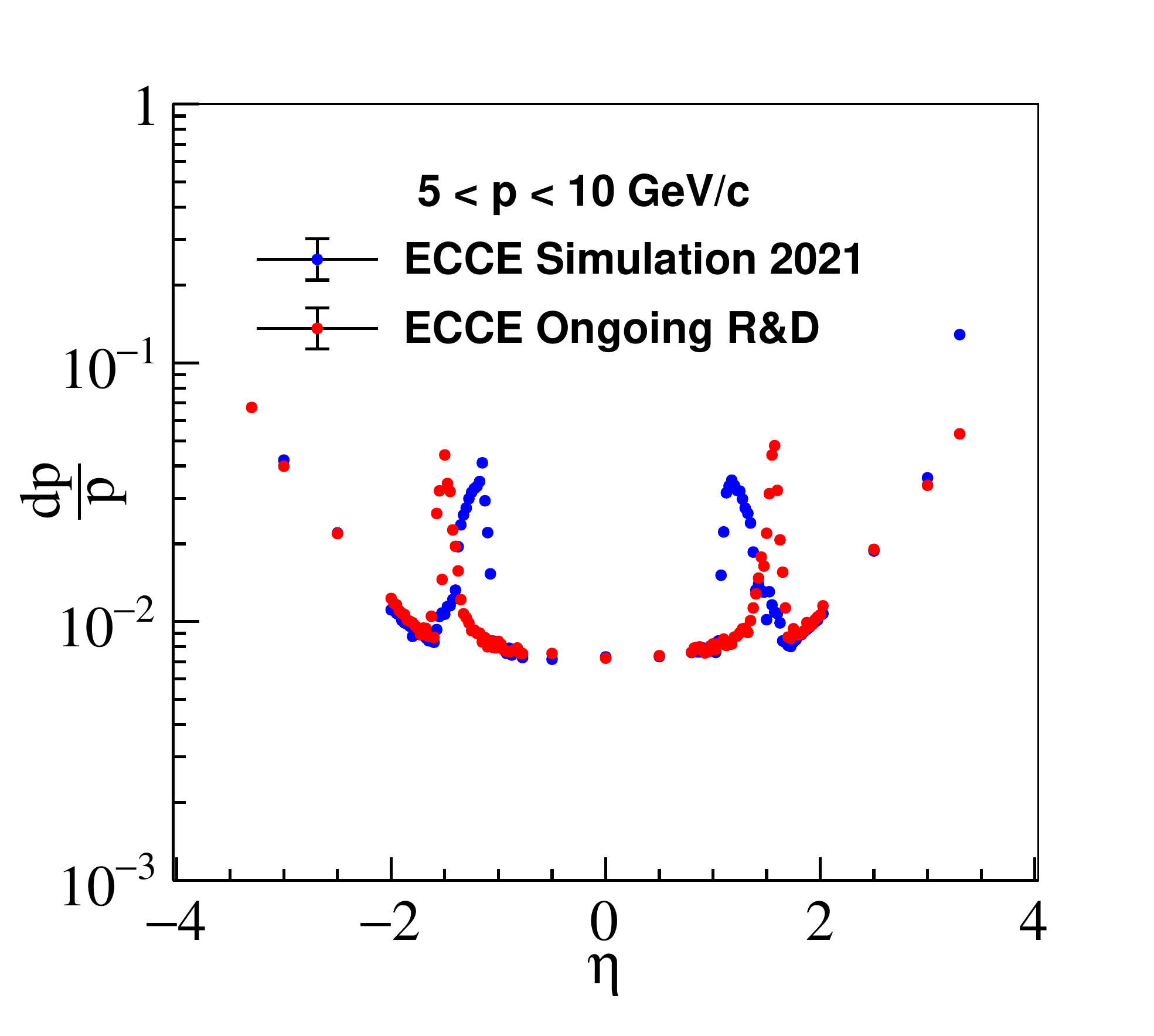}
    \includegraphics[width=0.3\textwidth]{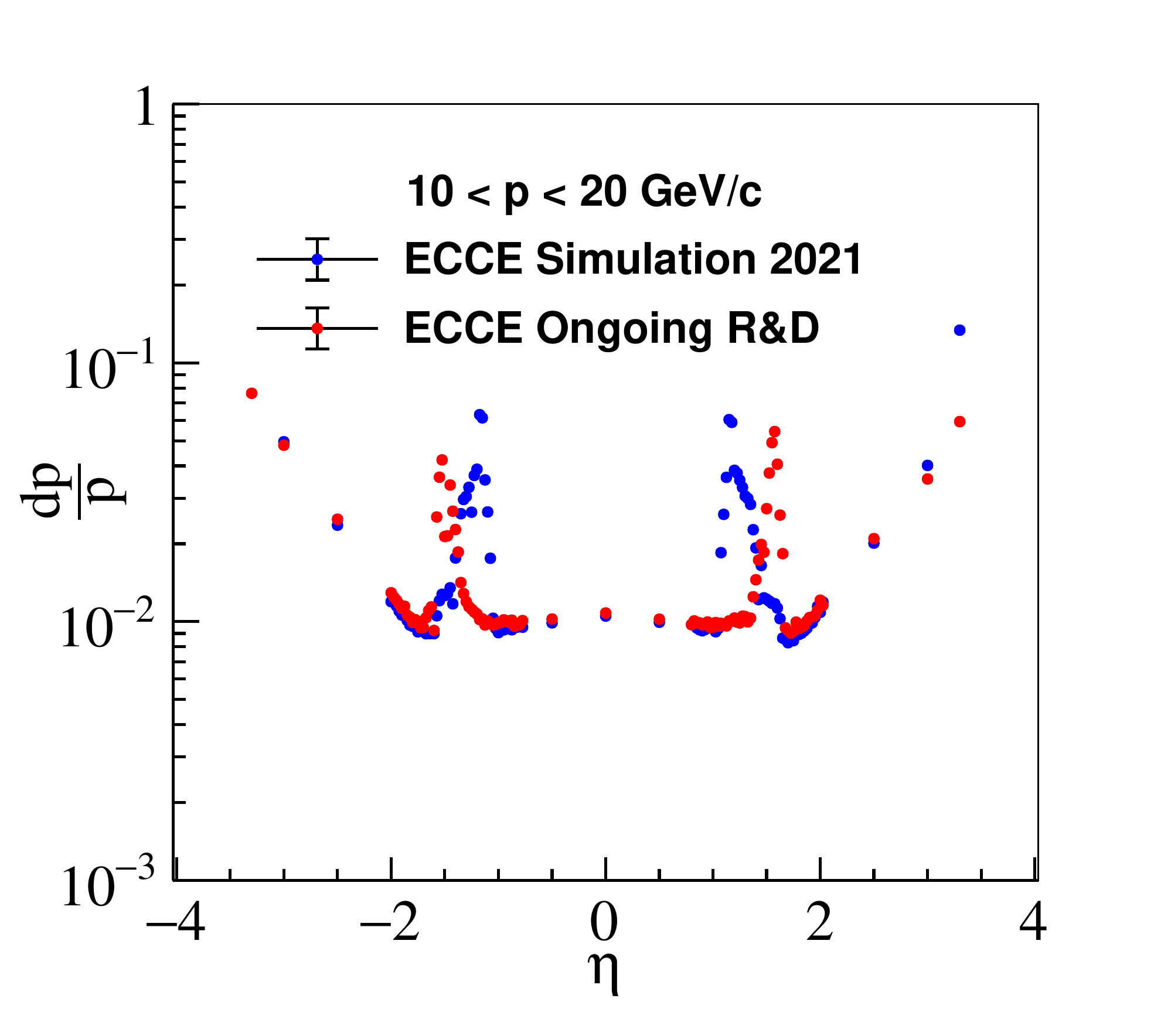}
    \caption{\textbf{Comparison between non-projective and projective inner tracker:}  a comparison in momentum resolution as a function of the pseudorapidity $\eta$ between the non-projective and the projective designs in three bins of momentum. The projective design concentrates the material in a smaller dead area resulting in better resolution on a wider range of the pseudorapidity. %Both designs have been assisted by AI. 
    %\cris{This is going to change into a row with multiple sub-figures (corresponding to bins in momentum); remember to reference this figure in the text.}
    }
    \label{fig:tracking_projective}
\end{figure*}

A comparison between the non-projective and projective designs of the inner tracker is also shown in Fig.~\ref{fig:tracking_projective}, where the projective design concentrates the material in a smaller dead area resulting in better resolution on a wider range of the pseudorapidity.
%

%\begin{comment}
%\end{comment}

\end{document}